\documentclass[aj_pt4]{aastex}
\usepackage{mathrsfs}
\usepackage{amsmath}
\usepackage{longtable}
\usepackage{booktabs}
\usepackage{hyperref}
\def\etal {et al.~}

\newbox\grsign \setbox\grsign=\hbox{$>$} \newdimen\grdimen \grdimen=\ht\grsign
\newbox\laxbox \newbox\gaxbox
\setbox\gaxbox=\hbox{\raise.5ex\hbox{$>$}\llap
     {\lower.5ex\hbox{$\sim$}}}\ht1=\grdimen\dp1=0pt
\setbox\laxbox=\hbox{\raise.5ex\hbox{$<$}\llap
     {\lower.5ex\hbox{$\sim$}}}\ht2=\grdimen\dp2=0pt

\shorttitle{CO OBSERVATION TOWARD HSVMT13}
\shortauthors{Yingjie \etal}

\newcommand{\co}{$^{12}$CO }                             
\newcommand{\xco}{$^{13}$CO }                            

\begin{document}

\title{Molecular Gas Toward the Gemini OB1 Molecular Cloud Complex II: CO Outflow Candidates with Possible WISE Associations}

\author{Yingjie Li \altaffilmark{1, 2}, Fa-Cheng Li \altaffilmark{1, 3}, Ye Xu \altaffilmark{1}, Chen  Wang \altaffilmark{1, 3}, Xin-Yu Du  \altaffilmark{1, 3}, Wenjin Yang \altaffilmark{1, 2}, Ji Yang \altaffilmark{1}}

\altaffiltext{1}{Purple Mountain Observatory, Chinese Academy of Sciences, Nanjing 210008, China, xuye@pmo.ac.cn}
\altaffiltext{2}{University of Science and Technology of China, Chinese Academy of Sciences, Hefei, Anhui 230026, China, liyj@pmo.ac.cn}
\altaffiltext{3}{Graduate University of the Chinese Academy of Sciences, 19A Yuquan Road, Shijingshan District, Beijing 100049, China}

\begin{abstract}
We present a large scale survey of CO outflows in the Gem OB1 molecular cloud complex and its surroundings using the Purple Mountain Observatory Delingha 13.7 m telescope. A total of 198 outflow candidates were identified over a large area ($\sim$ 58.5 square degrees), of which 193 are newly detected. Approximately 68\% (134/198) are associated with the Gem OB1 molecular cloud complex, including clouds GGMC 1, GGMC 2, BFS 52, GGMC 3 and GGMC 4. Other regions studied are: Local Arm (Local Lynds, West Front), Swallow, Horn, and Remote cloud. Outflow candidates in GGMC 1, BFS 52, and Swallow are mainly located at ring-like or filamentary structures. To avoid excessive uncertainty in distant regions ($\gtrsim 3.8$ kpc), we only estimated the physical parameters for clouds in the Gem OB1 molecular cloud complex and in the Local arm. In those clouds, the total kinetic energy and the energy injection rate of the identified outflow candidates are $\lesssim 1$\% and $\lesssim 3$\% of the turbulent energy and the turbulent dissipation rate of each cloud, indicating that the identified outflow candidates cannot provide enough energy to balance turbulence of their host cloud at the scale of the entire cloud (several pc to dozens of pc). The gravitational binding energy of each cloud is $\gtrsim 135$ times the total kinetic energy of the identified outflow candidates within the corresponding cloud, indicating that the identified outflow candidates cannot cause major disruptions to the integrity of their host cloud at the scale of the entire cloud.
\end{abstract}


\section{Introduction}

Outflows are intrinsic process during the early stage of the star formation, and are related to the mass-loss phase of stars of all masses \citep[e.g.,][and references therein]{ASG2007}. Outflowing supersonic winds can entrain and accelerate ambient gas and inject momentum and energy into the surrounding medium to produce an outflow, which in turn affects the dynamics and structure of its parent cloud \citep{ABG2010, NS1980}. Observational studies have revealed that outflows, even from low mass protostars, play an important role through their physical and chemical impact on their environments within a few parsecs \citep{FL2002, AS2006, ABG2010}. A detailed review for outflow feedback on parent cloud at the scale of 0.5--$10^2$ pc see \citet{FRC2014}.

In 1970s and 1980s, the early studies of the impact of outflows on surrounding gas were primarily in small fields ($\lesssim$ 10$\arcmin$), such as the cases of Orion KL \citep{KS1976}, L1551 \citep{SLP1980} and GL490 \citep{KS1976}. Later, in the 1980s, large-scale (unbiased) molecular outflow surveys were conducted of clouds with high-mass star formation activities, such as Mon OB1 \citep{ML1986} and the Orion southern cloud \citep{FST1986}\textbf{,} by using small (4--5 m) millimeter telescopes with beam size of $\sim 2\arcmin.5$. Owing to the poor performance of these telescopes, CO emission was diluted by the large beam sizes, and as such, only a few high-velocity outflows ($v > 10$ km s$^{-1}$) were detectable \citep{ABG2010}.

In recent years, some outflow surveys have been conducted of giant molecular clouds using search methods that cover different scales. Most studies search for outflows toward known star formation regions, covering scales of a few parsecs or less. While these methods see in great detail, they are necessarily limited in terms of providing a view of the outflow impact on the environments at larger scales. Outflow surveys by \citet{HFR2007} and \citet{HD2009} are examples. A method to study outflows at large scales is the unbiased search based on their dynamical properties, which can provide information to study the outflow impact on the entire cloud. Several studies have used this method to conduct large-scale molecular outflow surveys and have discussed outflow impact on their surroundings. Many large-scale unbiased outflow surveys \citep[e.g.,][]{ABG2010, NSB2012, LLQ2015} mainly focused on the famous star formation regions within $\sim 500$ pc to the solar system.

The outflows mentioned in the last paragraph are all powered by low-mass young stellar objects (YSOs). Fewer large scale (unbiased) molecular outflow surveys have been carried out on high mass star formation regions, let alone the study to investigate the impact of outflow activities on surrounding environments at large scales. The possible reasons are that they are relatively rare and far away from us \citep[see e.g.,][and reference therein]{ZFC2013, SBC2014, B2016}. The well-known and also the nearest outflow powered by high-mass YSOs is in the Orion BN/KL region with a distance of $\sim$ 414 pc \citep[][and references therein]{KS1976, B2016}. Most studies of this outflow have been carried out at the scale of less than 1 pc \cite[e.g.,][]{ZSH2009, PWF2009, PWZ2012}. This is insufficient to investigate their impact on surrounding environments at large scales.

It is worth to note, as it was indicated in \citet{BSG2002, BSS2002}, etc., that in most cases the detected outflows by single dish surveys (resolution angle larger than 5$\arcsec$) are actually the superposition of several outflows \citep{MSG2016}. For instance, the outflow identified by \citet{DME1998} with an angular resolution of $\sim $20$\arcsec$ is proved to be a composite of several outflows \citep{QZM2011} who observed with $\sim 3\arcsec$. Superposition is also found in interferometer researches, for example, a single outflow in the G31.41+0.31 \citep[][the angular resolutions are several 0.1$\arcsec$]{AHK2008, CBZ2011} is actually a double jet system \citep[][the angular resolution is $\sim$ 1 mas]{MLC2013}.

Despite all these cases, single dish surveys have the advantages in: 1) having wider field and mapping larger regions than interferometric instruments; 2) searching for (candidate bulk) outflows in much larger regions and therefore providing large samples for further detailed studies by interferometers; and 3) investigating the impact of (candidate bulk) outflows on their parent cloud at a large scale (i.e., few parsecs or larger). Therefore, large scale single dish surveys are still worth to be carried out.

\citet{WYX2017} studied the structures and physical properties of molecular (\co and its other two isotopic molecules) gas toward the Gem OB1 molecular cloud complex (Gem OB1 hereafter) and its surroundings (denote the whole observed region in this work as GOS for short hereafter). We have extended that study to a large-scale (unbiased) \co (frequently used outflow tracer) outflow survey of GOS. The distance of our study ranges from $\sim$ 400 pc to $\sim$ 8.7 kpc, which includes the Local arm, the Perseus arm, and may also include the Outer arm.

In Section 2, we describe the data used in this study, followed by the description of the Gem OB1 in the next section. We present our analysis and results in Section 4, and discussions is given in Section 5. Finally, the main results are summarized in Section 6.

\section{Data}

The observations of \co (1-0) (115.271 GHz) and \xco (1-0) (110.201 GHz) were used to search for outflows in GOS. These data are a part of the Milky Way Imaging Scroll Painting Project\footnotemark[4]\footnotetext[4]{See details in \url{http://www.radioast.nsdc.cn:81/english/shujukujieshao.php}.} (MWISP), which is a survey of the northern Galactic plane within $-10\degr.25\leq l\leq250\degr.25$ and $-5\degr.25\leq b\leq5\degr.25$, and other regions of interest. The $^{12}$CO and \xco molecular lines were observed simultaneously between September 2012 and May 2013 using the Purple Mountain Observatory Delingha (PMODLH) 13.7 m telescope with the 9-beam superconducting array receiver (SSAR), which works on the sideband separation mode and uses a fast Fourier transform spectrometer \citep{SYS2012}. The pointing accuracy was better than $4\arcsec$ during the observations. The receiver provides a total bandwidth of 1 GHz over 16,384 channels for both the upper-sideband and lower-sideband, corresponding to a velocity resolution of about 0.16 km s$^{-1}$ for $^{12}$CO, and 0.17 km s$^{-1}$ for $^{13}$CO, respectively.

The data were made in 234 cells of dimension $30\arcmin \times 30\arcmin$ and covered 58.5 square degrees ($186\degr.25\leq l\leq195\degr.25$, $-3\degr.75\leq b\leq2\degr.75$). The half-power beam-width (HPBW) was about 49$\arcsec$ for $^{12}$CO, and 51$\arcsec$ for $^{13}$CO. The typical system temperature during the observations was $\sim 280$ K for $^{12}$CO, and $\sim 185$ K for $^{13}$CO. The data were re-gridded to 30$\arcsec$ pixels. The main beam root mean squared noise (RMS) after main beam efficiency correction was $0.45$ K for $^{12}$CO, and $0.25$ K for $^{13}$CO, respectively.

\section{Overview of GOS}

The Gem OB1 was covered by the large-scale $^{12}$CO(1-0) survey of the entire Galactic plane and specific nearby clouds performed by the CfA 1.2 m telescope \citep{DUC1987, DHT2001}. However, it was hard to obtain detailed knowledge of the Gem OB1 due to the poor resolution of these surveys (i.e. $\sim 8\arcmin$). \citet{KOY1998} conducted a survey toward this region in \xco (1-0) using two 4 m telescopes at Nagoya University, and identified some star formation regions by using IRAS data. \citet{CSS1995a, CSS1995b} also conducted a survey of this region in both \co (1-0) and \xco (1-0) using the Five College Radio Astronomy Observatory (FCRAO) to study its star formation activities. However, outflow activities have not been studied yet.

GOS covers not only the Gem OB1, but also some other molecular clouds which are physically connected with each other in space. Paper I presented a detailed analysis of the composition of the molecular clouds in this region, which had greatly extended the studied region in \citet{BPV2009}. GOS can be decomposed to 10 sub-regions (all of them are molecular clouds, see details in Paper I). These \textbf{sub-regions} are listed in Table 1 (part of Table 1 in Paper I) that summarizes the distances and the main component velocity ranges ($^{13}$CO). The molecular clouds GGMC 1, GGMC 2, BFS 52, GGMC 3 and GGMC 4 compose a molecular cloud complex: the Gem OB1. This complex is located in the Perseus arm at a distance of $\sim$2 kpc and is associated with HII regions SH247, SH252, SH254--258, and BFS 52. The Local Lynds dark clouds and the West Front clouds are located in the Local arm. The mean spectra of these 10 sub-regions are presented in Figure 1, including the $^{12}$CO and \xco lines.

\begin{deluxetable}{lccl}
\centering
\tablecolumns{4}
\tabletypesize{\normalsize}
\tablewidth{0pt}
\tablecaption{Distances and Velocity Ranges ($^{13}$CO) of the 10 Sub-regions}
\tablehead{
 \colhead{Sub-region} & \colhead{Distance} & \colhead{Velocity Range} & \colhead{Reference} \\
 \colhead{}   & \colhead{(kpc)} & \colhead{(km s$^{-1}$)}  & \colhead{}
}
\startdata
GGMC 1       & 2.0 & [0, 10]  & 1 \\
GGMC 2       & 2.0 & [-2, 16] & 2, 3, 4 \\
BFS 52       & 2.0 & [-2, 16] & 2, 3, 4 \\
GGMC 3       & 2.0 & [-2, 16] & 2, 3, 4 \\
GGMC 4       & 2.0 & [-4, 14] & 2, 3, 4 \\
Local Lynds & 0.4 & [-5, 7] & 5, 6 \\
West Front  & 0.6 & [-5, 8] & 7 \\
Swallow     & 3.8? & [11, 19] & 8 \\
Horn        & 3.8? & [11, 20] & 8 \\
Remote      & 8.7? & [16, 28] & 8, 9 \\
\enddata
\tablewidth{15cm}
\tablerefs{(1) \citet{CSS1995a}; (2) \citet{RMB2009}; (3) \citet{NNH2011}; (4) \citet{RBR2010}; (5) \citet{BM1974}; (6) \citet{EMP2013}; (7) \citet{PAA2015}; (8) Paper I; (9) \citet{MJF1979}.}
\tablecomments{? = the distances are undetermined, we adopted the recommended distances in Paper I\textbf{.}}
\end{deluxetable}

\begin{figure}
\centering
\includegraphics[angle=-90,width=0.9\textwidth]{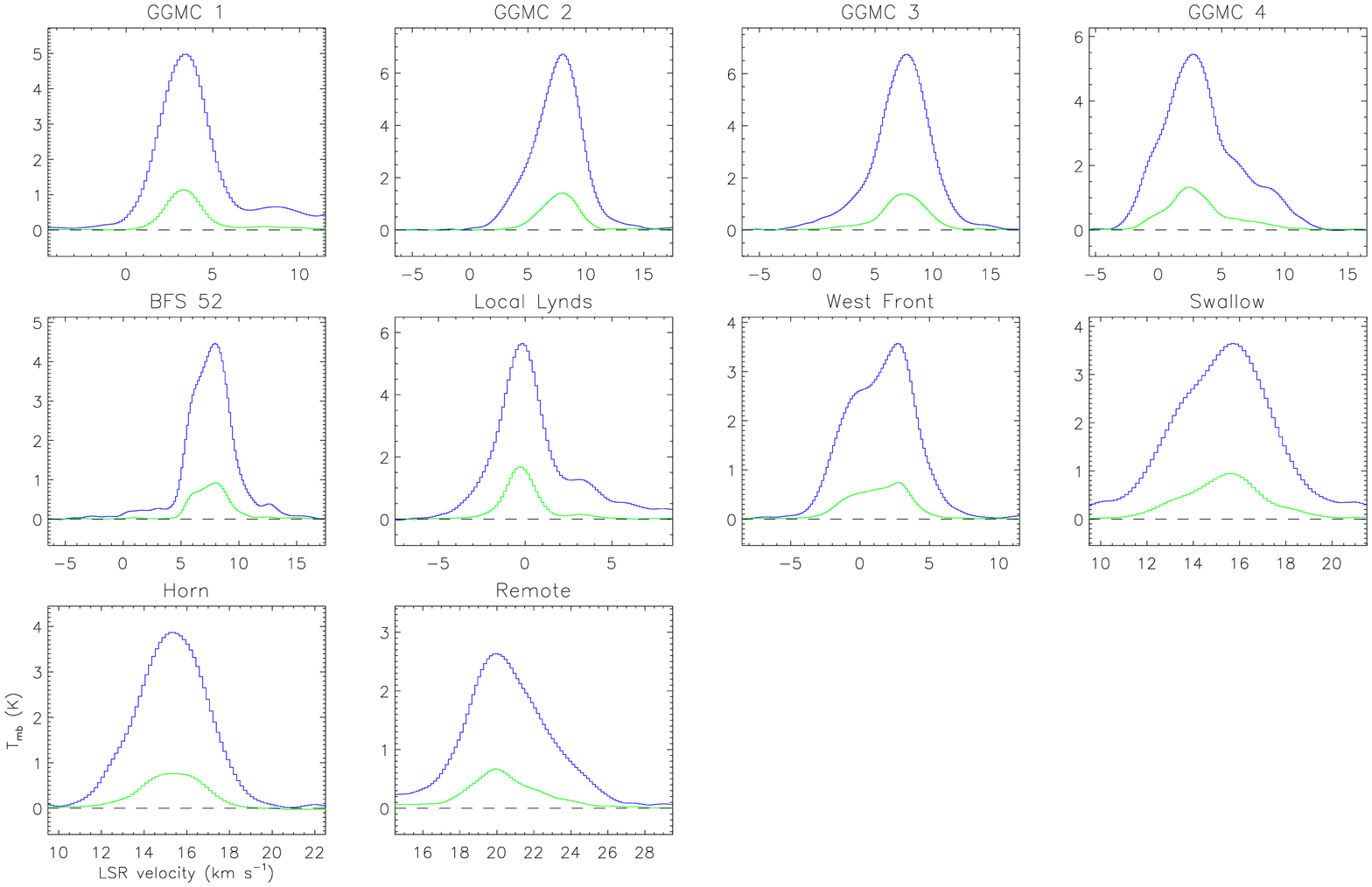}
\caption{Mean spectra of the 10 sub-regions. Blue is $^{12}$CO and green is $^{13}$CO. The regions that contribute to the mean spectra are those where the \xco emission is larger than 3$\times$RMS.}
\end{figure}

\section{Data Analysis and Result}

\subsection{Outflow Identification}

We conducted a large-scale molecular outflow survey of GOS, based on a set of semi-automated IDL scripts\footnotemark[5].\footnotetext[5]{\url{https://github.com/AlphaLFC/gemini_outflows}.} we searched for and identified outflows using three-dimensional \co data (based on the longitude-latitude-velocity space), and traced the cores found in the three-dimensional \xco data. We searched for the minimum velocity extents from the velocity characteristics of the three-dimensional contour diagram, conducted line diagnoses of positions with the minimum velocity extents, and finally obtained the outflow candidate samples in different quality level. The specific method is presented in Appendix A. Table 2 shows the results of the first five outflow candidates in GGMC 2. The details of each outflow candidate in each sub-region are shown Appendix C (table). We did not include outflow candidates with quality level D in Table 2 and Appendix C, because they were removed from the final sample (see Appendix A. 6, a scoring system with quality level A-D was introduced to classify all outflow candidates, where A denoted the most likely outflow candidates, and D denoted the abandoned outflow candidates).

\begin{deluxetable}{crrccccc}
\centering
\setlength\tabcolsep{3pt}
\tablecolumns{8}
\tabletypesize{\small}
\tablewidth{0pt}
\tablecaption{Examples of Outflow candidates in GGMC 2}
\tablehead{
 \colhead{Index} & \colhead{$l$} & \colhead{$b$} & \colhead{Blue Line Wing} & \colhead{Red Line Wing} & \colhead{Quality level} & \colhead{WISE Emission} & \colhead{New Detection} \\
 \colhead{} & \colhead{(\degr)} & \colhead{(\degr)} & \colhead{(km s$^{-1}$)} & \colhead{(km s$^{-1}$)} & \colhead{Blue$|$Red} & \colhead{} & \colhead{}
}
\startdata
G2-1 & 192.455 &  0.005 & $(1.2, 3.5)$   & \nodata          & B        & Y? & Y  \\
G2-2 & 192.580 &  0.210 & \nodata        & $(10.2, 12.4)$   & B        & Y? & Y  \\
G2-3 & 192.598 & -0.048 & $(-0.2, 4.3)$  & $(11.0, 15.9)$   & A$|$A    & Y  & N  \\
G2-4 & 192.608 & -0.129 & $(0.5, 4.3)$   & \nodata          & C        & Y? & Y  \\
G2-5 & 192.655 & -0.210 & \nodata        & $(11.0, 13.0)$   & A        & Y? & Y
\enddata
\vspace{-11pt}
\tablecomments{Y = Yes, Y? = possible, N = Not a new detection. The quality level see Appendix A. 6. Since the outflow candidates with score D were removed from the final sample, we do not include them in this table.}
\end{deluxetable}

Table 3 summarizes the outflow survey results where outflow candidates with quality level D are not included. 134 outflow candidates are located in Gem OB1 which is located in the Perseus arm, 96\% being newly detected. Twenty-one detected outflow candidates are associated with the Local arm, which includes Local Lynds and West Front. Forty-three outflow candidates were detected beyond the Perseus arm.

\begin{deluxetable}{lccccc}
\centering
\setlength\tabcolsep{3pt}
\tablecolumns{6}
\tabletypesize{\small}
\tablewidth{0pt}
\tablecaption{Outflow candidate Sample Statistics}
\tablehead{
 \colhead{Sub-region} & \colhead{Numbers of Outflow} & \colhead{Numbers of} & \multicolumn{3}{c}{Quality Level}\\
 \cline{4-6}
 \colhead{} & \colhead{Total(Blue$|$Red)} & \colhead{Bipolar Outflow} & \colhead{A} & \colhead{B} & \colhead{C}
}
\startdata
GGMC 1      & 34(22$|$18)    & 6  &  3$|$8\%   & 19$|$48\%  & 18$|$45\% \\
GGMC 2      & 20(11$|$16)    & 7  &  7$|$26\%  & 13$|$48\%  &  7$|$26\% \\
BFS 52       & 17(12$|$13)    & 8  &  8$|$32\%  & 10$|$40\%  &  7$|$28\% \\
GGMC 3      & 45(28$|$31)    & 14 & 25$|$42\%  & 28$|$47\%  &  6$|$10\% \\
GGMC 4      & 18(13$|$8)     & 3  &  8$|$38\%  &  8$|$38\%  &  5$|$24\% \\
Local Lynds & 2(2$|$1)       & 1  &  1$|$33\%  &  2$|$67\%  &  0$|$0\%  \\
West Front  & 19(13$|$11)    & 5  &  1$|$4\%   & 13$|$54\%  & 10$|$42\% \\
Swallow     & 10(5$|$7)      & 2  &  4$|$33\%  &  7$|$58\%  &  1$|$8\%  \\
Horn        & 9(8$|$9)       & 8  &  7$|$41\%  &  8$|$47\%  &  2$|$12\% \\
Remote      & 24(17$|$16)    & 9  &  9$|$27\%  & 15$|$45\%  &  9$|$27\% \\
Total       & 198(131$|$130) & 63 & 73$|$28\%  & 123$|$47\% & 65$|$25\% \\
\enddata
\vspace{-11pt}
\tablecomments{In ``Quality Level'' column, we presented the numbers and percentages and separated them with a ``$|$''. The definition of quality level see Appendix A. 6.}
\end{deluxetable}

There are six previously known outflows in GOS. Five of them are included in the detected outflow candidates, the other which are associated with CB39 \citep[][]{YC1992, WEZ1995} was manually removed from the outflow sample, because of the quality level (see Figure A. 4, and Appendix A. 6). The two works shared a similar HPBW ($\sim 48\arcsec$) and sensitivity (0.25 K per 250 kHz channel, corresponding to a velocity resolution of 0.65 km s$^{-1}$ at 115 GHz) to this work. However, they did not show an image of the outflow, we were unable to figure out the reason why this one was failed to be detected.

\subsection{The Limitation of the Survey}

Although we found a large number of outflow candidates, some with faint emission, low velocity, complex environment or small size could be missed. In addition, owing to the moderate resolution, the identified outflow candidates, especially for those at large distances, can in fact be multi-outflows. We reported specifically the limitation of this survey from three aspects in the following. First, the sensitivity is 0.45 K per 61 kHz channel (corresponding to $\sim$ 0.16 km s$^{-1}$ at 115.271 GHz) for $^{12}$CO, some fainter or low-velocity outflows could be missed. Second, the interaction between outflows and ambient gas at large scales makes their morphologies uncertain, which would present confused morphologies in the maps. Therefore, many outflow candidates detected with one lobe (either red or blue) does not indicate that they are genuine monopolars. Their counterflows can be too faint/confused to be detected.

Last and most important, owing to moderate resolution, small size outflows ($\lesssim$ 49$\arcsec$) cannot be detected, and the identified outflow candidates could be multi-outflows. Two CO (1-0) outflows surveys with the similar HPBW ($\sim$ 46$\arcsec$) were selected to estimate the possible multi-outflows. The two outflow surveys were toward Perseus and Taurus molecular clouds \citep[hereafter PTC,][]{ABG2010, LLQ2015}\footnotemark[6].\footnotetext[6]{Sensitivities from PTC (i.e., $\sim$ 0.43 K per 0.06 km s$^{-1}$ channel and $\sim$ 0.28 K per 0.26 km s$^{-1}$ channel, respectively) were superior to the current study. In addition, PTC are closer, therefore some faint or low-velocity outflows can be detected.} The mean outflow radius in PTC were $\sim$ 0.5 and $\sim$ 0.3 pc \citep{ABG2010, LLQ2015}, corresponding to $\gtrsim 154\arcsec$ and $\gtrsim 103\arcsec$ at 400--600 pc, respectively. These angular sizes are greatly larger than the HPBW of the current study ($\sim$ 49$\arcsec$), indicating that most outflow candidates in Local Lynds and West Front are resolvable. The mean angular outflows size in PTC corresponded to $\sim$ 31$\arcsec$ and $\sim$ 52$\arcsec$ at 2 kpc, respectively. These angular sizes are close to the HPBW, which indicates that some outflow candidates in Gem OB1 are bulk outflow candidates. Therefore, we expected a bias towards higher values of the derived quantities but lower detection rates. However, the mean angular outflows size in PTC corresponded to $\lesssim$ 22$\arcsec$ at $\geq$ 3.8 kpc, which is significantly less than the HPBW. Outflow candidates were highly clumpy in regions with distance $\geq$ 3.8 kpc.

What is more, the distances in Swallow, Horn or Remote are undetermined, see Paper I. To avoid the great uncertainties in physical parameters toward these three sub-regions, we did not investigate the outflow candidates' properties and only studied the spatial distribution of outflow candidates in these regions in the following contents, see Section 5.1. The reserved information, especially the positions and line wing velocity ranges of outflow candidates presented in Appendix C (table), will contribute to future high-resolution investigations.

As a pilot outflow survey, this study probably underestimates outflow impact by mis-identifying outflow candidates. Higher resolution observations would be followed up to identify individual outflows.

\subsection{The Amount of Possible Outflow Emission Missed}

Outflows column density, $N_\mathrm{o}$, i.e., the number of outflows per square pc, can indicate the level of undetected outflows. Because YSO column density, $N_\mathrm{Y}$, in PTC \citep[$\sim$ 1.3 pc$^{-2}$,][]{EDJ2009, ABG2010, RPM2010} is slightly higher than that in the part of Gem OB1 that surrounds SH 252 \citep[i.e., 0.5 pc$^{-2}$,][]{JPO2012, JPS2013}, indicating that star formation in Gem OB1 is less active than that in PTC. We therefore expected a smaller $N_\mathrm{o}$ in Gem OB1 (located in the Perseus arm) and the Local arm, where star formation in the Local arm is generally less active than that in the Perseus arm. \citet{ABG2010} and \citet{LLQ2015} found $N_\mathrm{o}\approx$ 0.2 pc$^{-2}$ in PTC. Given the same distance ($\sim$ 2 kpc) and a similar $N_\mathrm{Y}$ \citep[0.6 pc$^{-2}$,][]{KAG2008} to Gem OB1, \citet{GBW2011} only found 40 CO (2-1) outflows in W5 within $\sim$ 4060 pc$^{2}$, where the HPBW ($\sim$ 14$\arcsec$) and sensitivity (0.1 K per 0.4 km s$^{-1}$ channel) were superior to the current study. Combining the surveys in PTC and in W5, the actual number of outflows should be greater than the detected count by factors of no more than 16, 10, 8, 13, 11, 3, and 13 (i.e., scale-up factors, SUFs) for the first seven sub-regions in Table 1, respectively.

\subsection{Physical Parameters}

The estimated physical parameters included candidate outflow lobe velocity ($\langle\Delta v_{\mathrm{lobe}}\rangle$), length ($l_{\mathrm{lobe}}$), mass ($M_{\mathrm{lobe}}$), momentum ($P_{\mathrm{lobe}}$), kinetic energy ($E_{\mathrm{lobe}}$), dynamical timescale ($t_\mathrm{lobe}$), and luminosity ($L_{\mathrm{lobe}}$). The calculations were described in Appendix B. As an example, Table 4 lists the physical properties of the first five outflow candidates in GGMC 2, and the complete sample are cataloged in Appendix D (for the definition of the location of outflow candidates, see Appendix A. 5.).

\begin{deluxetable}{llrrrrrrrrr}
\centering
\setlength\tabcolsep{3pt}
\tablecolumns{11}
\tabletypesize{\small}
\tablewidth{0pt}
\tablecaption{Physical Properties of the Outflow Candidate Samples}
\tablehead{
\colhead{Index} & \colhead{Lobe} & \colhead{$l$} & \colhead{$b$} & \colhead{$\langle\Delta v_{\mathrm{lobe}}\rangle$} & \colhead{$l_{\mathrm{lobe}}$} & \colhead{$M_{\mathrm{lobe}}$} & \colhead{$P_{\mathrm{lobe}}$} & \colhead{$E_{\mathrm{lobe}}$} & \colhead{$t_\mathrm{lobe}$} & \colhead{$L_{\mathrm{lobe}}$} \\
 \colhead{} & \colhead{} & \colhead{($\degr$)} &  \colhead{($\degr$)} & \colhead{(km s$^{-1}$)} & \colhead{(pc)} & \colhead{(M$_{\odot}$)} & \colhead{(M$_{\odot}$ km s$^{-1}$)} & \colhead{(10$^{43}$ erg)} & \colhead{(10$^5$ yr)} & \colhead{(10$^{30}$ erg s$^{-1}$)}
}
\startdata
G2-1     &   Blue &  192.455 &  0.005   & 3.0  &  2.75   &     0.99   &    2.99   &     8.97      &   6.8     &     4.2    \\
G2-2     &   Red &  192.580 &  0.210   &  1.9  & 1.34   &     0.65   &    1.24   &     2.32      &   4.1     &     1.8    \\
G2-3     &   Blue &  192.604 &  -0.054  & 5.4  &  1.49   &     1.19   &    6.44   &     34.50     &   2.0     &     53.9   \\
         &   Red &  192.589 &  -0.042  &  5.6  & 2.10   &     3.97   &    22.30  &     124.00    &   2.4      &     164.0  \\
G2-4     &   Blue &  192.608 &  -0.129  & 4.8  &  1.49   &     0.92   &    4.39   &     20.80     &   2.4     &     28.4   \\
G2-5     &   Red &  192.655 &  -0.210  &  4.2  & 1.83   &     0.84   &    3.53   &     14.70     &   3.4     &     14.1
\enddata
\end{deluxetable}

Three major issues can affect the uncertainties of the derived outflow parameters, namely, inclination, blending, and opacity \citep{AG2001B}. The inclination is the angle between the long axis of the outflow and the line of sight. Assuming a random distribution of inclination angles, the mean value is given by $\langle\theta\rangle=\int^{\pi/2}_{0}\theta \sin \theta d\theta=57\degr.3$ \citep{BATC1996, DAM2014}. Then, the velocity, length, and dynamical timescale should be multiplied by a factor of about 1.9, 1.2, and 0.6, respectively. When we considered that some low-velocity outflowing gas blend into ambient gas, the outflow mass should be scaled up by a factor of $\sim$2.0, as previous studies have shown \citep{ML1985, ABG2010, NSB2012}. Caution was taken that those corrections were only applied to statistics. We did not estimate the correction for optical depth, which was discussed in Appendix B. Table 5 summarizes the correction factors, which also include corrections for outflow momentum, kinetic energy and luminosity. The following discussions is based on the physical parameters after correcting for inclination and blending. Table 6 shows the global physical properties (after correction) of outflow candidates in each sub-region.

\begin{deluxetable}{ccccc}
\centering
\tablecolumns{5}
\tabletypesize{\small}
\tablewidth{0pt}
\tablecaption{Correction Factors}
\tablehead{
 \colhead{Quantity} & \colhead{Inclination} & \colhead{Inclination} & \colhead{Blending} & \colhead{Total} \\
 \colhead{} & \colhead{Dependence} & \colhead{Correction} & \colhead{Correction} & \colhead{Correction}
}
\startdata
$\langle\Delta v_{\mathrm{lobe}}\rangle$ & 1/$\cos \theta$ & 1.9 & \nodata & 1.9 \\
$l_{\mathrm{lobe}}$ & 1/$\sin \theta$ & 1.2 & \nodata & 1.2 \\
$M_{\mathrm{lobe}}$ & \nodata & \nodata & 2.0 & 2.0 \\
$P_{\mathrm{lobe}}$ & 1/$\cos \theta$ & 1.9 & 2.0 & 3.8 \\
$E_{\mathrm{lobe}}$ & 1/$\cos^2 \theta$ & 3.4 & 2.0 & 6.8 \\
$t_{\mathrm{lobe}}$ & $\cos \theta/ \sin \theta$ & 0.6 & \nodata & 0.6 \\
$L_{\mathrm{lobe}}$ & $\sin \theta/\cos^3 \theta$ & 5.3 & 2.0 & 10.6 \\
\enddata
\tablewidth{15cm}
\vspace{-11pt}
\tablecomments{The physical quantities include outflow lobe's velocity ($\langle\Delta v_{\mathrm{lobe}}\rangle$), length ($l_{\mathrm{lobe}}$), mass ($M_{\mathrm{lobe}}$), momentum ($P_{\mathrm{lobe}}$), kinetic energy ($E_{\mathrm{lobe}}$), the dynamical timescale ($t_\mathrm{lobe}$), and luminosity ($L_{\mathrm{lobe}}$).}
\end{deluxetable}

\begin{deluxetable}{lrrrrrr}
\centering
\setlength\tabcolsep{3pt}
\tablecolumns{7}
\tabletypesize{\footnotesize}
\tablewidth{0pt}
\tablecaption{Global Outflow Candidate Properties (corrected) in Each Sub-region}
\tablehead{
 \colhead{Sub-region} & \colhead{Velocity} & \colhead{Mass} & \colhead{Momentum} & \colhead{Kinetic Energy} & \colhead{Luminosity} & \colhead{Dynamical Timescale} \\
 \colhead{} & \colhead{(km s$^{-1}$)} & \colhead{(M$_{\odot}$)} & \colhead{(M$_{\odot}$ km s$^{-1}$)} & \colhead{(10$^{44}$ erg)} & \colhead{(10$^{32}$ erg s$^{-1}$)} & \colhead{(10$^5$ yr)}
}
\startdata
GGMC 1      & 4.6 & 66.8     & 308.6   &  139.4   & 11.9   & 3.9  \\
GGMC 2      & 6.5 & 108.6    & 608.2   &  457.6   & 69.2   & 3.2  \\
BFS 52       & 3.9 & 34.8     & 144.4   &  61.2    & 6.0    & 3.9  \\
GGMC 3      & 6.8 & 184.8    & 1394.6  &  1122.0  & 159.5  & 2.8  \\
GGMC 4      & 7.2 & 48.6     & 471.2   &  474.6   & 113.0  & 2.8  \\
Local Lynds & 6.0 & 0.2      & 1.1     &  0.7     & 0.8    & 0.7  \\
West Front  & 5.8 & 4.8      & 28.5    &  18.4    & 6.7    & 1.1  \\
\enddata
\vspace{-11pt}
\tablecomments{The physical parameter in the second and last column is the mean value of all lobes within each sub-region.}
\end{deluxetable}

\section{Discussion}

\subsection{The Spatial Distribution of Outflow candidates}

Figures 2--11 map the outflow candidate distributions for the 10 sub-regions in Table 1 in order, and HII regions are marked. Then we inspected the locations of outflow candidates by eye in the context of \xco gas emission.

\begin{figure}[!ht]
\centering
\includegraphics[height=\textwidth,angle=-90]{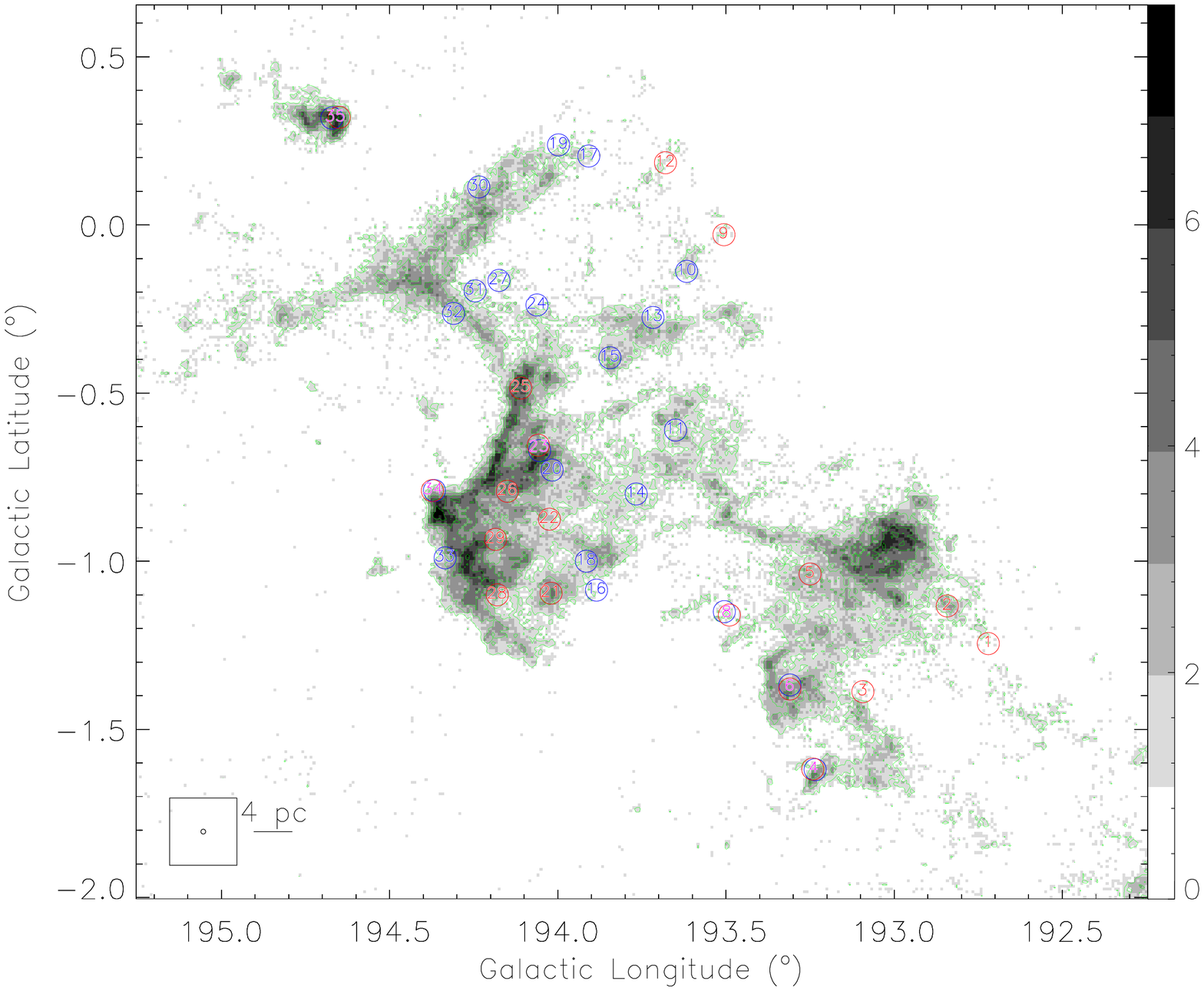}
\caption{Outflow candidate distribution in GGMC 1. The background gray-scale map and contours are the integrated intensity map of $^{13}$CO in the range of [2, 9] km s$^{-1}$, and the color bar is in units of K km s$^{-1}$. The contours start at 5$\sigma$ with $\sigma=0.27$ K km s$^{-1}$, and the contour interval is 20\% of the difference between the peak intensity and 5$\sigma$. The blue and red open circles denote the blue and red lobes, respectively. The blue/red numbers in the circles are the indexes of blue/red lobes, and the magenta numbers mark the bipolar outflow candidates. The definition of the location of outflow candidates refers to Appendix A. 5. The beam of the \xco observations (the open circle in the insert) and physical scale bar are reported in the bottom left of the panel.}
\label{fig:map_GGMC1}
\end{figure}

\begin{figure}[!ht]
\centering
\includegraphics[height=\textwidth,angle=-90]{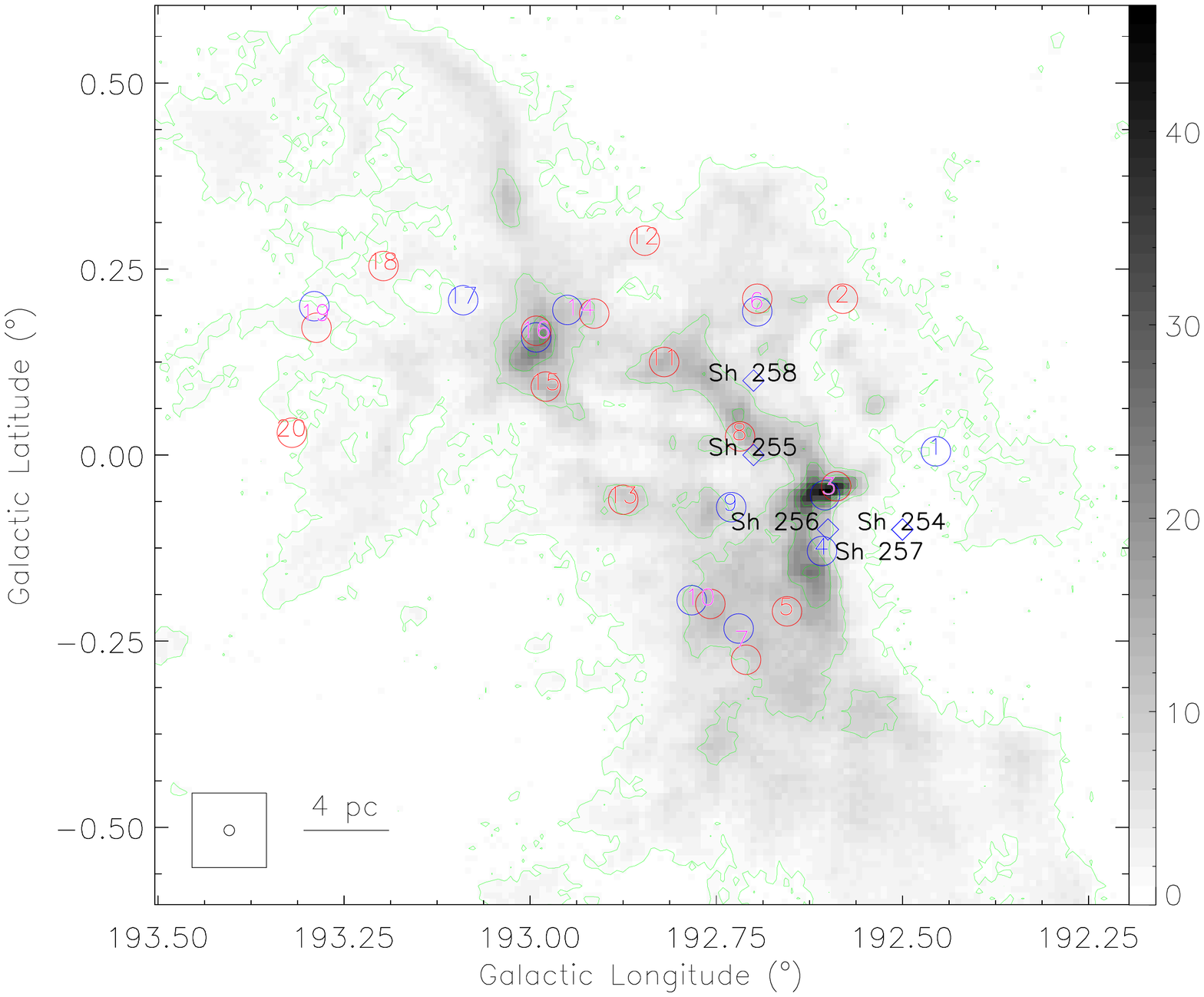}
\caption{Outflow candidate distribution in GGMC 2. The background gray-scale map and contours are the integrated intensity map of $^{13}$CO in the range of [5, 11] km s$^{-1}$, and the color bar is in units of K km s$^{-1}$. The contour levels are 5, 30, 60, 90 120 $\times$ 0.25 K km s$^{-1}$ (1$\sigma$). The blue and red open circles denote the blue and red lobes, respectively. The blue/red numbers in the circles are the indexes of blue/red lobes, and the magenta numbers mark the bipolar outflow candidates. The definition of the location of outflow candidates refers to Appendix A. 5. Diamonds represent HII regions Sh 254--258 \citep{S1959}. The beam of the \xco observations (the open circle in the insert) and physical scale bar are reported in the bottom left of the panel.}
\label{fig:map_GGMC2}
\end{figure}

\begin{figure}[!ht]
\centering
\includegraphics[height=\textwidth,angle=-90]{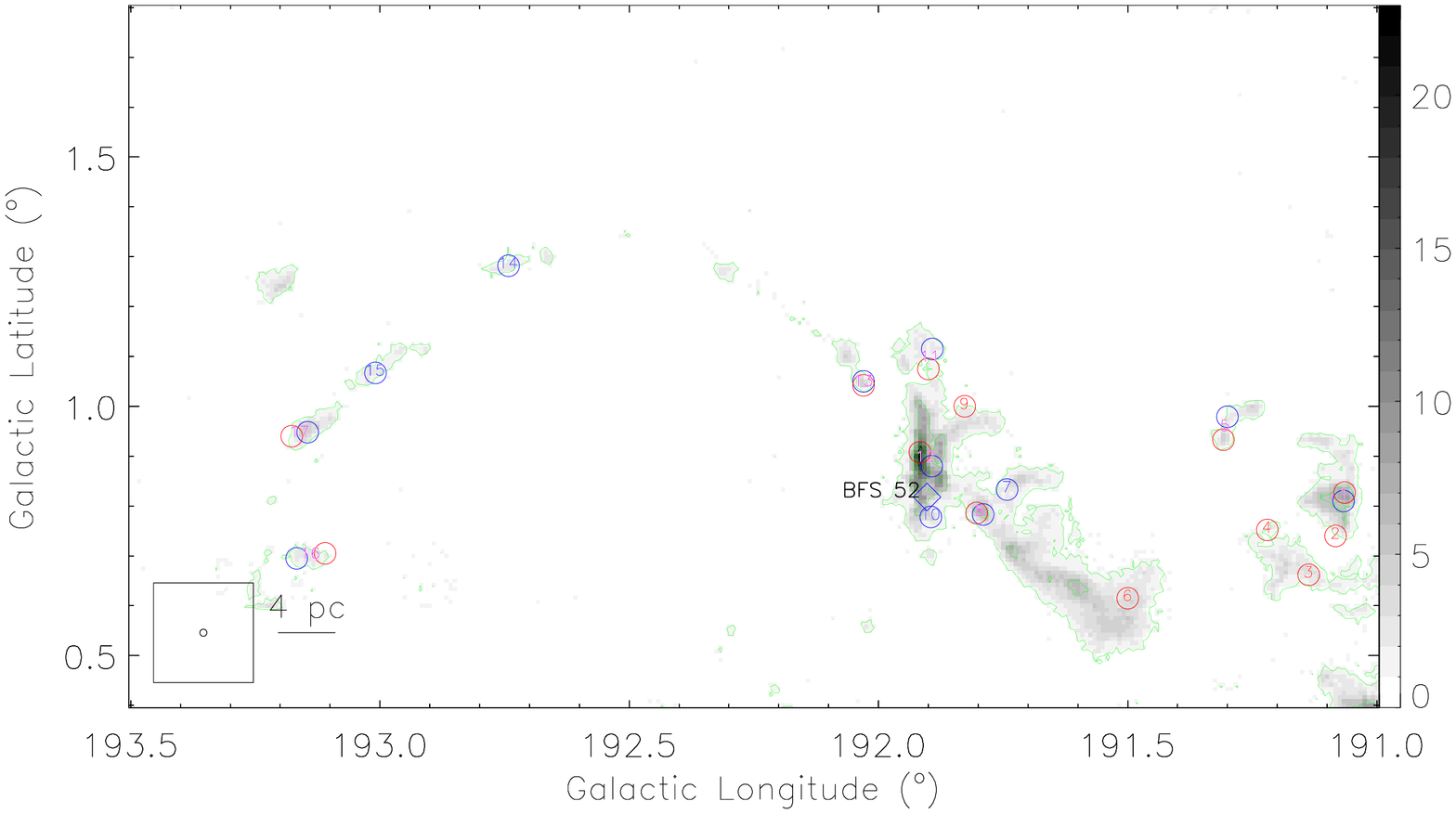}
\caption{Outflow candidate distribution in BFS 52. The background gray-scale map and contours are the integrated intensity map of $^{13}$CO in the range of [5, 11] km s$^{-1}$, and the color bar is in units of K km s$^{-1}$. The contours start at 5$\sigma$ with $\sigma=0.25$ K km s$^{-1}$, and the contour interval is 20\% of the difference between the peak intensity and 5$\sigma$. The blue and red open circles denote the blue and red lobes, respectively. The blue/red numbers in the circles are the indexes of blue/red lobes, and the magenta numbers mark the bipolar outflow candidates. The definition of the location of outflow candidates refers to Appendix A. 5. Diamond represents HII region BFS 52 \citep{BFS1982, CSS1995b}. The beam of the \xco observations (the open circle in the insert) and physical scale bar are reported in the bottom left of the panel.}
\label{fig:map_BFS 52}
\end{figure}

\begin{figure}[!ht]
\centering
\includegraphics[height=\textwidth,angle=-90]{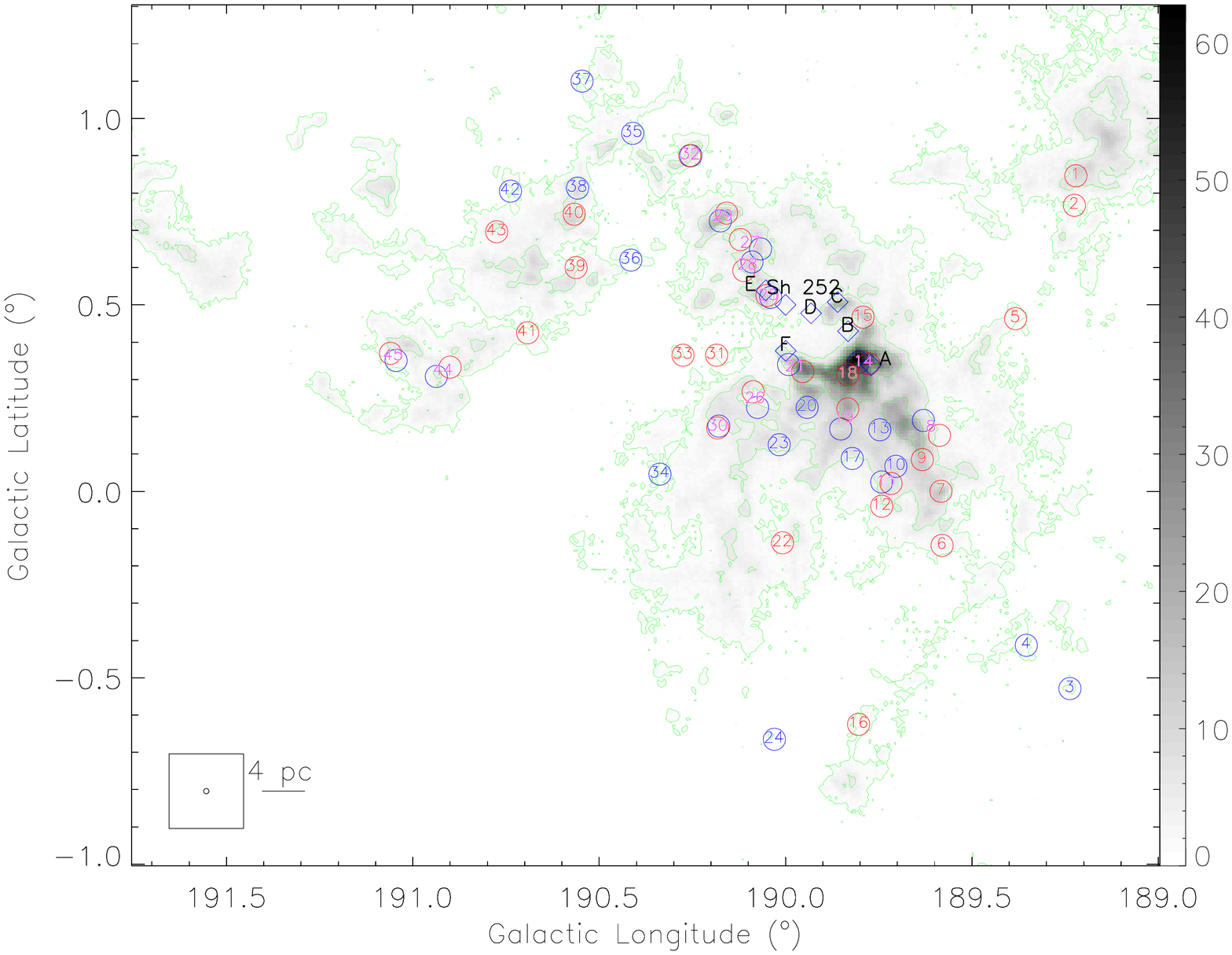}
\caption{Outflow candidate distribution in GGMC 3. The background gray-scale map and contours are the integrated intensity map of $^{13}$CO in the range of [5, 11] km s$^{-1}$, and the color bar is in units of K km s$^{-1}$. The contour levels are 5, 20, 60, 100, 140 $\times$ 0.25 K km s$^{-1}$ (1$\sigma$). The blue and red open circles denote the blue and red lobes, respectively. The blue/red numbers in the circles are the indexes of blue/red lobes, and the magenta numbers mark the bipolar outflow candidates. The definition of the location of outflow candidates refers to Appendix A. 5. Diamonds represent HII regions Sh 252 \citep{S1959}, and A, B, C, D, E, F in \citet{FHI1977}. The beam of the \xco observations (the open circle in the insert) and physical scale bar are reported in the bottom left of the panel.}
\label{fig:map_GGMC3}
\end{figure}

\begin{figure}[!ht]
\centering
\includegraphics[height=\textwidth, angle=-90]{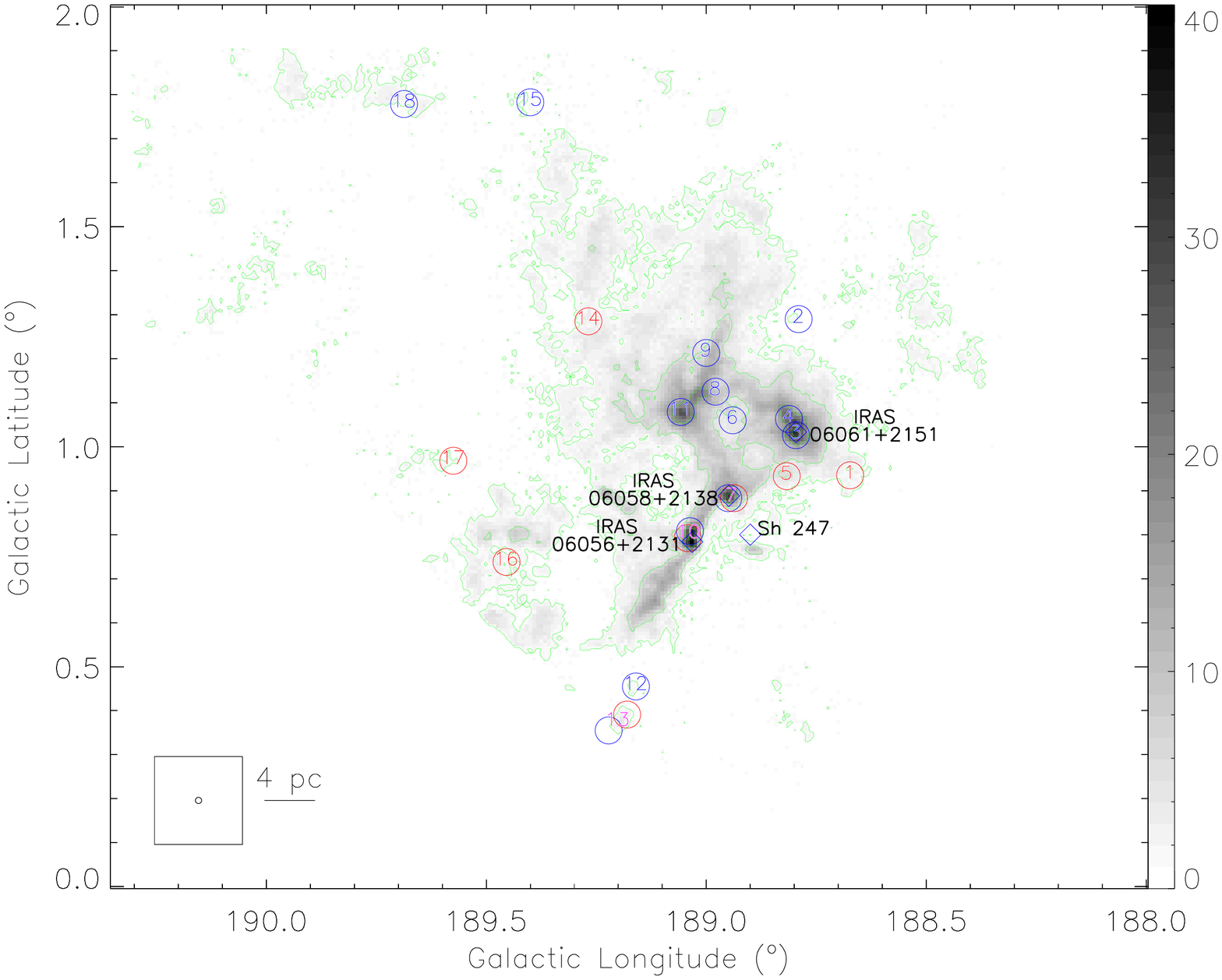}
\caption{Outflow candidate distribution in GGMC 4. The background gray-scale map and contours are the integrated intensity map of $^{13}$CO in the range of [-3, 5] km s$^{-1}$, and the color bar is in units of K km s$^{-1}$. The contour levels are 5, 20, 60, 100, 140 $\times$ 0.29 K km s$^{-1}$ (1$\sigma$). The blue and red open circles denote the blue and red lobes, respectively. The blue/red numbers in the circles are the indexes of blue/red lobes, and the magenta numbers mark the bipolar outflow candidates. The definition of the location of outflow candidates refers to Appendix A. 5. Diamonds represent HII regions Sh 247 \citep{S1959}, and another three compact HII regions \citep[IRAS 06051+2151, IRAS 06058+2138 and IRAS 06065+2131 in][]{GIK2000}. The beam of the \xco observations (the open circle in the insert) and physical scale bar are reported in the bottom left of the panel.}
\label{fig:map_GGMC4}
\end{figure}

\begin{figure}[!ht]
\centering
\includegraphics[height=\textwidth, angle=-90]{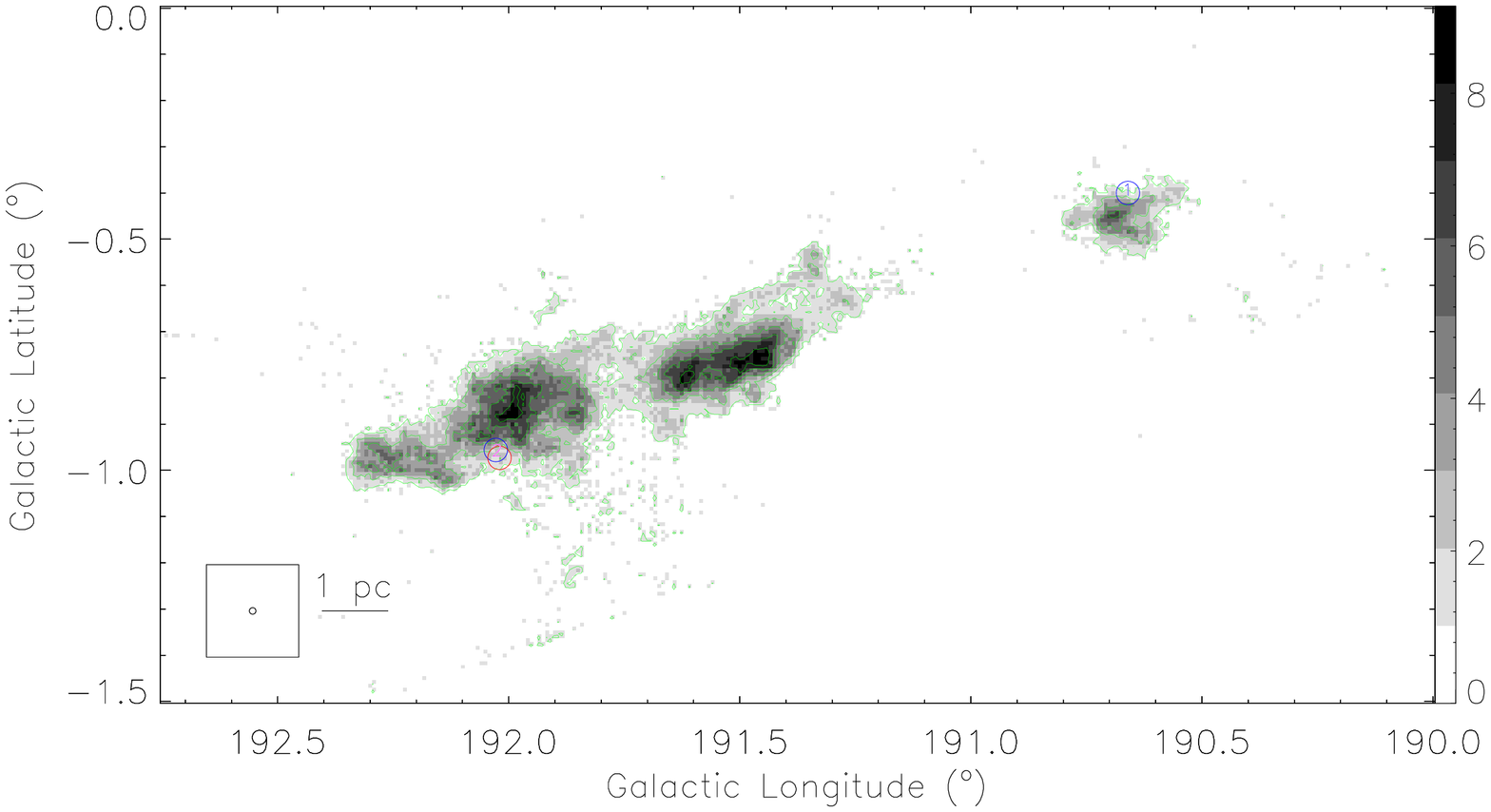}
\caption{Outflow candidate distribution in Local Lynds. The background gray-scale map and contours are the integrated intensity map of $^{13}$CO in the range of [5, 11] km s$^{-1}$, and the color bar is in units of K km s$^{-1}$. The contours start at 5$\sigma$ with $\sigma=0.25$ K km s$^{-1}$, and the contour interval is 20\% of the difference between the peak intensity and 5$\sigma$. The blue and red open circles denote the blue and red lobes, respectively. The blue/red numbers in the circles are the indexes of blue/red lobes, and the magenta numbers mark the bipolar outflow candidates. The definition of the location of outflow candidates refers to Appendix A. 5. The beam of the \xco observations (the open circle in the insert) and physical scale bar are reported in the bottom left of the panel.}
\label{fig:map_lynds}
\end{figure}

\begin{figure}[!ht]
\centering
\includegraphics[height=0.8\textheight,angle=-90]{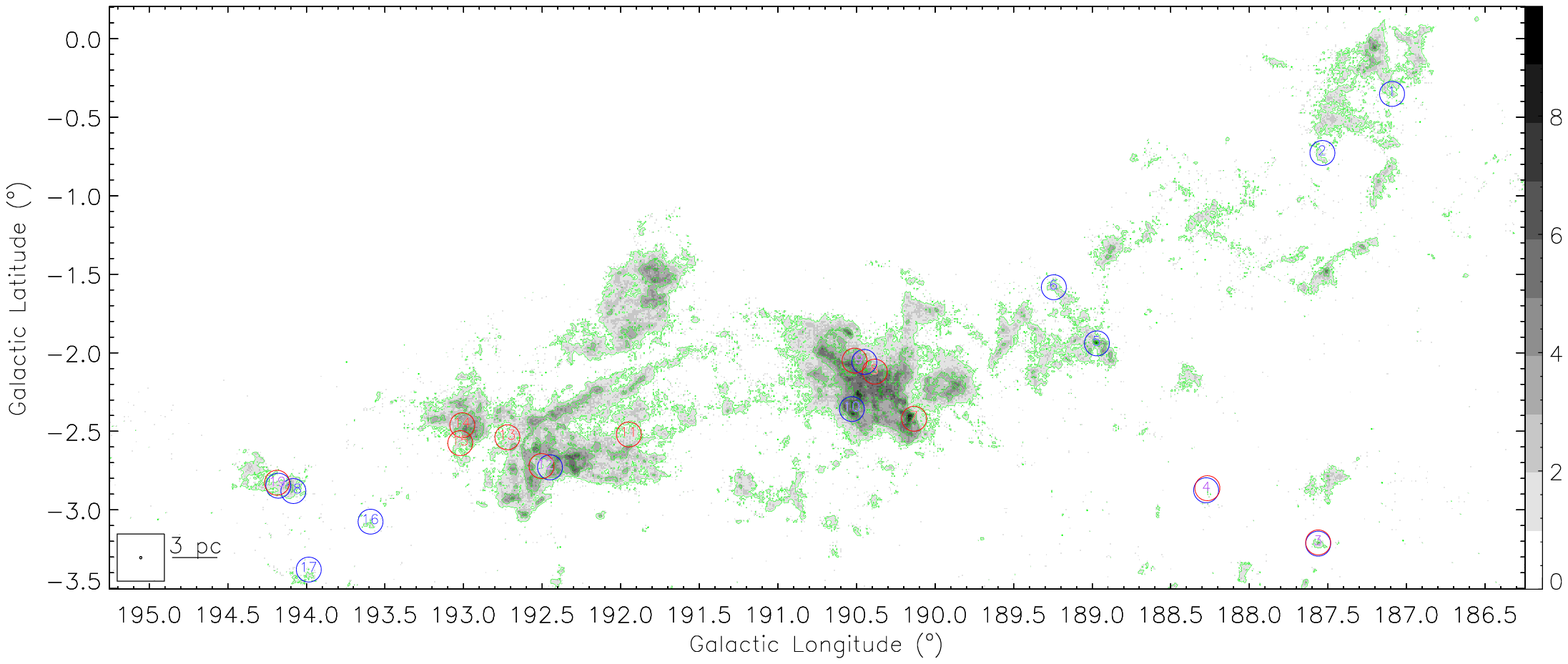}
\caption{Outflow candidate distribution in West Front. The background gray-scale map and contours are the integrated intensity map of $^{13}$CO in the range of [-1, 4] km s$^{-1}$, and the color bar is in units of K km s$^{-1}$. The contours start at 5$\sigma$ with $\sigma=0.23$ K km s$^{-1}$, and the contour interval is 20\% of the difference between the peak intensity and 5$\sigma$. The blue and red open circles denote the blue and red lobes, respectively. The blue/red numbers in the circles are the indexes of blue/red lobes, and the magenta numbers mark the bipolar outflow candidates. The definition of the location of outflow candidates refers to Appendix A. 5. The beam of the \xco observations (the open circle in the insert) and physical scale bar are reported in the bottom left of the panel.}
\label{fig:map_west}
\end{figure}

\begin{figure}[!ht]
\centering
\includegraphics[height=0.7\textheight]{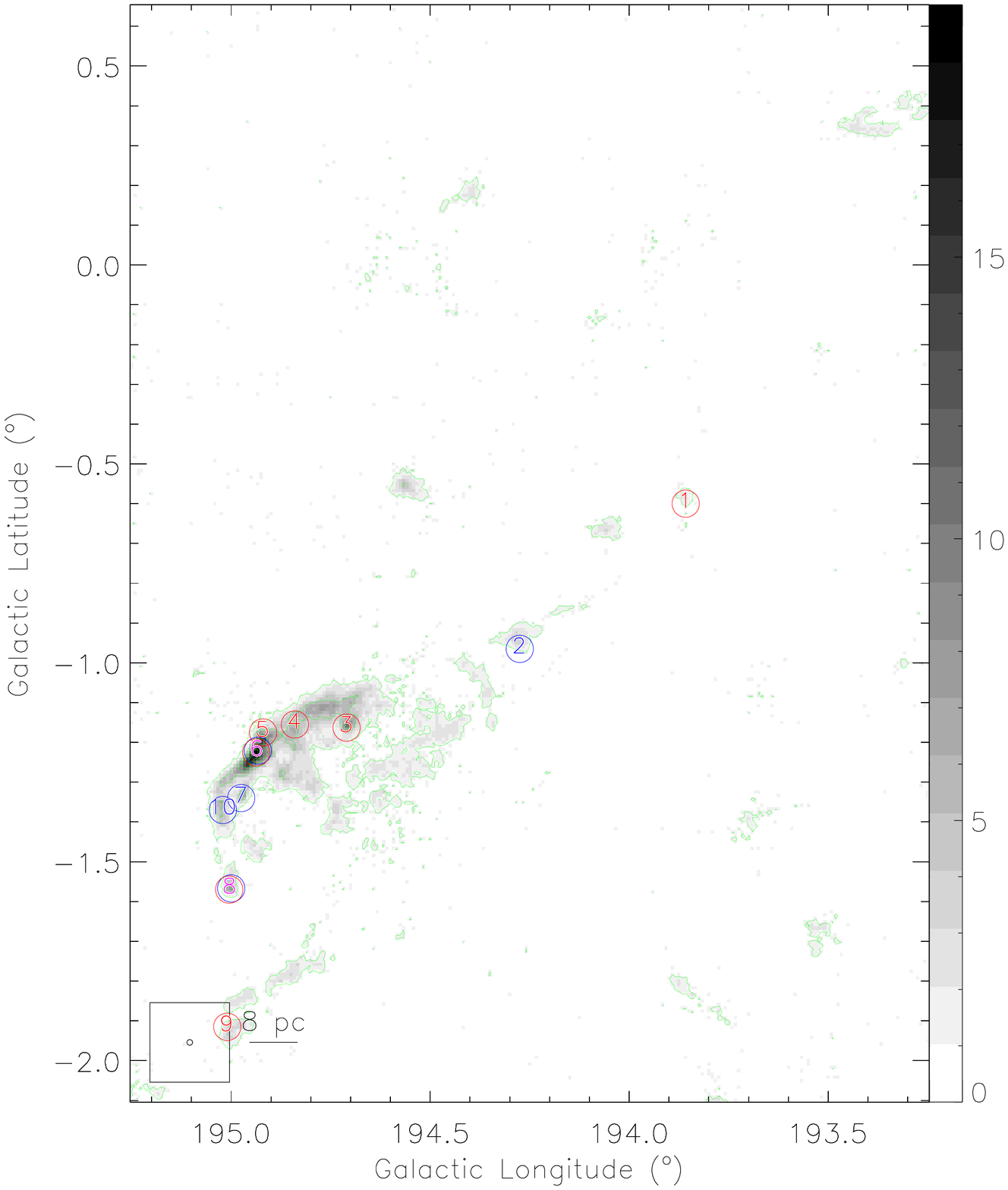}
\caption{Outflow candidate distribution in Swallow. The background gray-scale map and contours are the integrated intensity map of $^{13}$CO in the range of [5, 11] km s$^{-1}$, and the color bar is in units of K km s$^{-1}$. The contours start at 5$\sigma$ with $\sigma=0.25$ K km s$^{-1}$, and the contour interval is 20\% of the difference between the peak intensity and 5$\sigma$. The blue and red open circles denote the blue and red lobes, respectively. The blue/red numbers in the circles are the indexes of blue/red lobes, and the magenta numbers mark the bipolar outflow candidates. The definition of the location of outflow candidates refers to Appendix A. 5. The beam of the \xco observations (the open circle in the insert) and physical scale bar are reported in the bottom left of the panel.}
\label{fig:map_swallow}
\end{figure}

\begin{figure}[!ht]
\centering
\includegraphics[height=0.7\textheight,angle=-90]{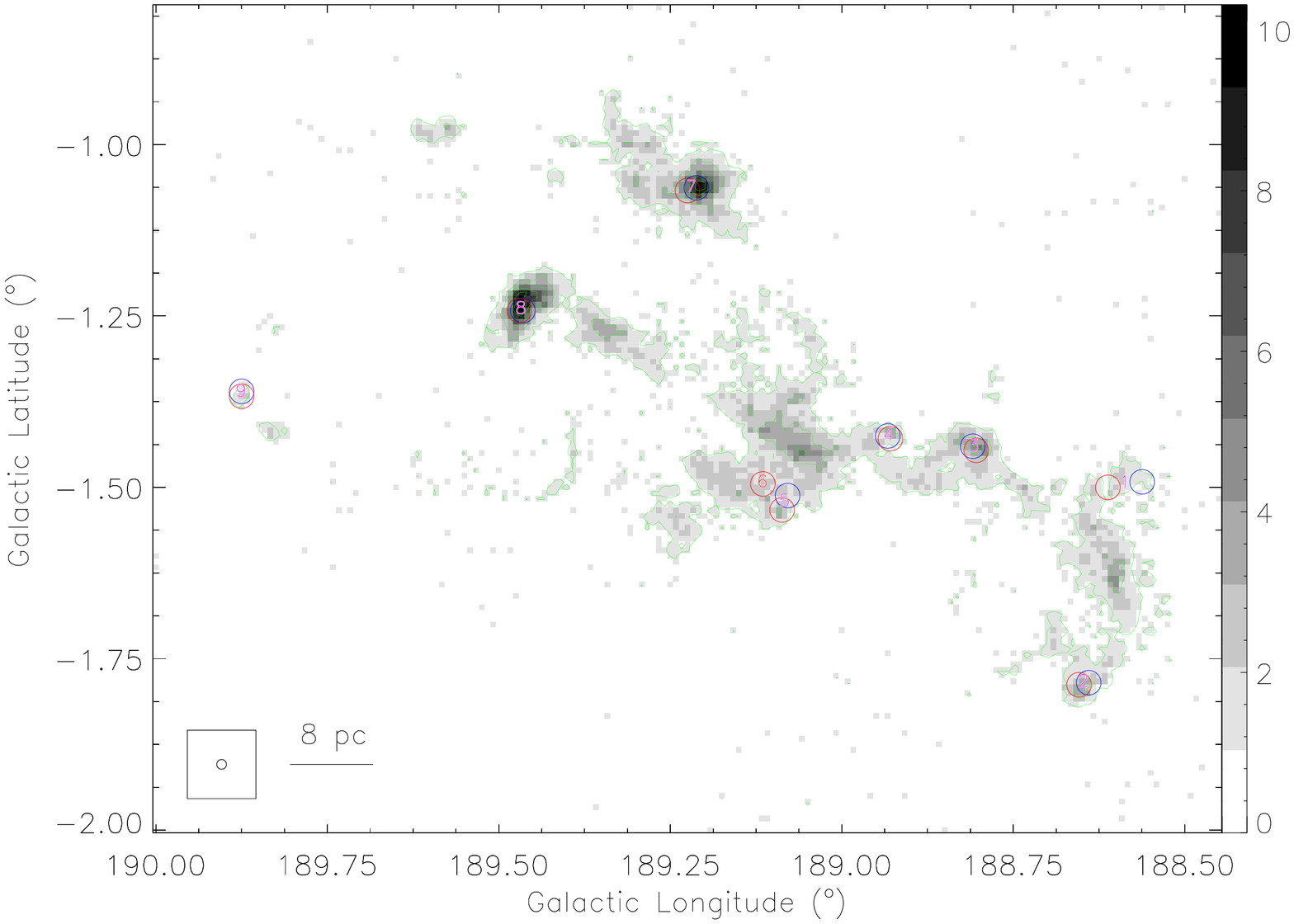}
\caption{Outflow candidate distribution in Horn. The background gray-scale map and contours are the integrated intensity map of $^{13}$CO in the range of [12, 18] km s$^{-1}$, and the color bar is in units of K km s$^{-1}$. The contours start at 5$\sigma$ with $\sigma=0.25$ K km s$^{-1}$, and the contour interval is 20\% of the difference between the peak intensity and 5$\sigma$. The blue and red open circles denote the blue and red lobes, respectively. The blue/red numbers in the circles are the indexes of blue/red lobes, and the magenta numbers mark the bipolar outflow candidates. The definition of the location of outflow candidates refers to Appendix A. 5. The beam of the \xco observations (the open circle in the insert) and physical scale bar are reported in the bottom left of the panel.}
\label{fig:map_horn}
\end{figure}

\begin{figure}[!ht]
\centering
\includegraphics[height=0.7\textheight,angle=-90]{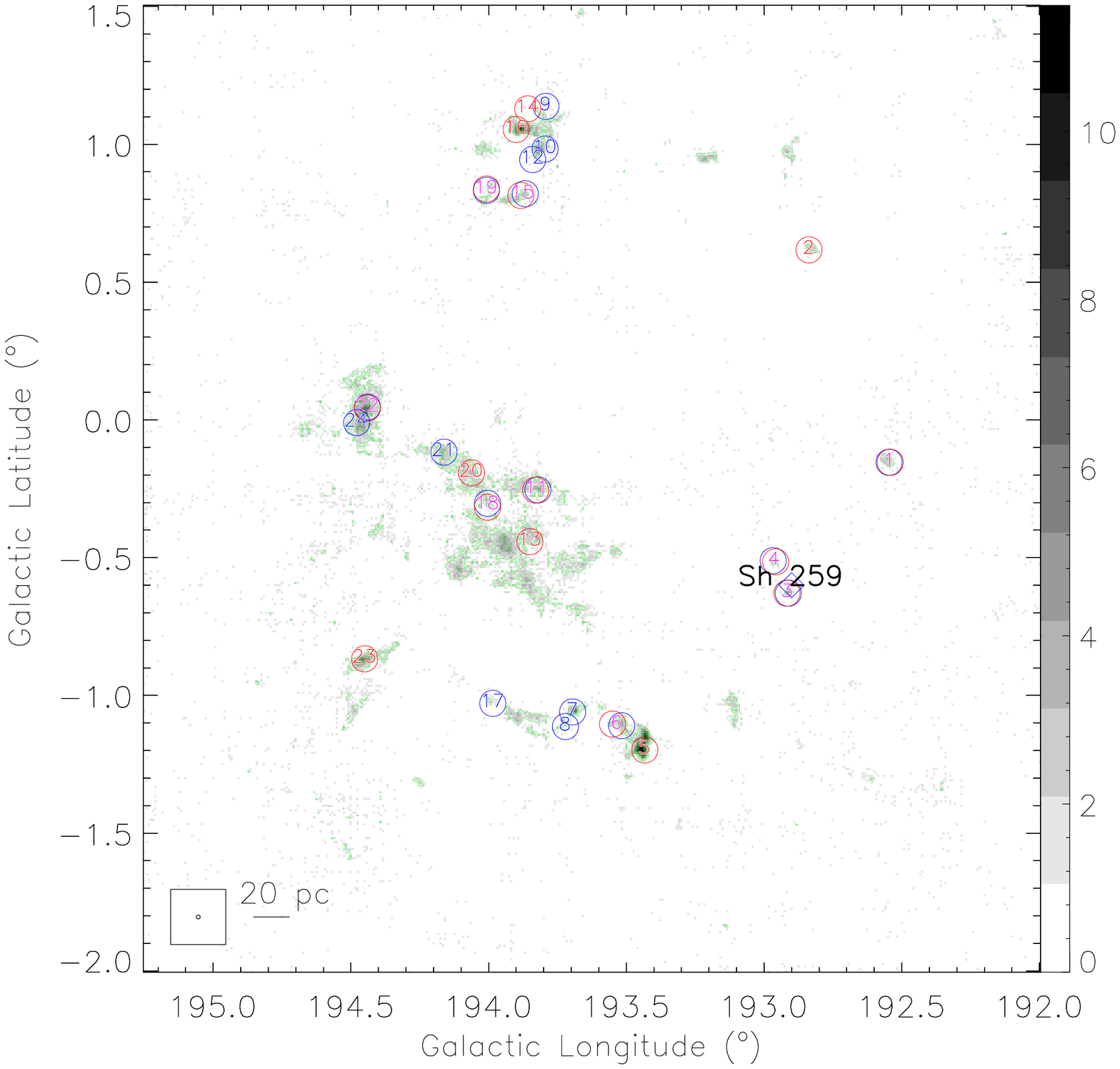}
\caption{Outflow candidate distribution in Remote. The background gray-scale map and contours are the integrated intensity map of $^{13}$CO in the range of [18, 28] km s$^{-1}$, and the color bar is in units of K km s$^{-1}$. The contours start at 5$\sigma$ with $\sigma=0.33$ K km s$^{-1}$, and the contour interval is 20\% of the difference between the peak intensity and 5$\sigma$. The blue and red open circles denote the blue and red lobes, respectively. The blue/red numbers in the circles are the indexes of blue/red lobes, and the magenta numbers mark the bipolar outflow candidates. The definition of the location of outflow candidates refers to Appendix A. 5. Diamond represents HII region Sh 259 \citep{S1959}. The beam of the \xco observations (the open circle in the insert) and physical scale bar are reported in the bottom left of the panel.}
\label{fig:map_remote}
\end{figure}

Generally, outflow candidates in GGMC 1 (Figure 2),  BFS 52 (Figure 4) and Swallow (Figure 9) are mainly associated with \xco intensity peaks, and are preferentially located at ring-like or filamentary structures (see Paper I figures, marked by ``A'' and ``F'', respectively). However, outflow candidates are ubiquitously distributed in GGMC 2 (Figure 3), GGMC 3 (Figure 5), GGMC 4 (Figure 6), Local Lynds (Figure 7), West Front (Figure 8), Horn (Figure 10) and Remote (Figure 11).

Table 7 summarizes the strong outflow candidates (who simultaneously have the top 25\% values in each sub-region in velocity, momentum, energy, and luminosity) and their distributions\footnotemark[7]\footnotetext[7]{In some cases, when one physical parameter (e.g., velocity) of an outflow candidate is among the top 25\% in a sub-region, the other physical parameters are not in the top 25\% in the sub-region. Therefore, the total number of strong outflow candidates is lower than 25\% of the total number of outflow candidates in each sub-region.}. To better describe the outflow candidates distributions, we classified them into four groups as follows: group I is located in filamentary or ring-like structures; group II is close to \xco intensity peaks (its distance from the local highest contour boundary is no more than 3$\arcmin$); group III is close to HII regions (the nearest HII regions is no more than 5$\arcmin$ away); group IV is deviated from the \xco intensity peaks (opposite to group II). We did not consider outflow candidates in Local Lynds because there were only two outflow candidates in this region, or in Swallow, Horn or Remote, because their physical parameters were not calculated. Some strong outflow candidates in group I, and most of them are associated with \xco intensity peaks (group II). The majority of the strong outflow candidates are in group II, and most of them are close to HII regions (group III) indicating that star formation was somewhat relative with HII regions. However, a few \textbf{strong} outflow candidates deviated from the \xco intensity peaks (group IV).

\begin{deluxetable}{lcccccc}
\centering
\setlength\tabcolsep{4pt}
\tablecolumns{7}
\tabletypesize{\scriptsize}
\tablewidth{0pt}
\tablecaption{The Strong Outflow Candidates and Their Distributions}
\tablehead{
 \colhead{Sub-region} & \colhead{Amount\tablenotemark{a}} & \colhead{Indexes } & \colhead{Group I} & \colhead{Group II} & \colhead{Group III} & \colhead{Group IV}
}
\startdata
GGMC 1       & 3 & G1-21, G1-22 G1-29 & \nodata & G1-21 & \nodata & G1-22, G1-29 \\
GGMC 2       & 1 & G2-3 & \nodata & G2-3 & G2-3\tablenotemark{b} & \nodata  \\
BFS 52       & 2 & B-3, B-12 & B-3, B-12 & B-3, B-12 & B-12\tablenotemark{c} & \nodata \\
GGMC 3       & 4 & G3-14, G3-17, G3-18, G3-21 & \nodata & G3-14, G3-18, G3-21 & G3-14\tablenotemark{d}, G3-18\tablenotemark{d}, G3-21\tablenotemark{e} & G3-17 \\
GGMC 4      & 3 & G4-7, G4-8, G4-10 & G4-10 & G4-7, G4-10 & G4-7\tablenotemark{f}, G4-10\tablenotemark{g} & G4-8  \\
West Front  & 2 & W-7, W-9 & \nodata & W-7 & W-9 & \nodata  \\
Total       & 15 & \nodata & \nodata & \nodata & \nodata & \nodata \\
\enddata
\vspace{-11pt}
\tablenotetext{a}{The total number of strong outflow candidates.}
\tablenotetext{b}{It is close to HII regions Sh 256 and Sh 257 \citep{S1959}.}
\tablenotetext{c}{It is close to HII region BFS 52 \citep{BFS1982, CSS1995b}.}
\tablenotetext{d}{It is close to HII region A \citep{FHI1977}.}
\tablenotetext{e}{It is close to HII region F \citep{FHI1977}.}
\tablenotetext{f}{It is close to HII region IRAS 06058+2138 \citep{GIK2000}.}
\tablenotetext{g}{It is close to HII region IRAS 06056+2131 \citep{GIK2000}.}
\end{deluxetable}

\subsection{Feedback of Outflow Candidates}

In this section, we will discuss the feedback from outflow candidates. Table 8 shows the physical properties of each sub-region. Most clouds in Gem OB1, including GGMC 2, BFS 52, GGMC 3, GGMC 4, are massive star formation regions (Paper I). For massive molecular outflows, although the derived physical parameters include contributions from multi-outflows, they are most likely dominated by one massive flow \citep{BSS2002, TM2002}. It is possible that, in most cases, the discovered outflow candidates are clustered outflow candidate systems as shown in \citet{BSS2002} and \citet{TM2002}. We neglect the effects due to multiple sources in the following. Future high resolution study will be of great interest.

\subsubsection{Impact of Outflow Candidates---Energy}

We analyzed the impact of the total energy of outflow candidates on the whole cloud. The turbulent energy of the cloud is given by \citep{ABG2010}:
\begin{equation}
 E_{\mathrm{turb}}=0.5M_{\mathrm{cloud}}\sigma^2_{\mathrm{3d}},
\end{equation}
where $M_{\mathrm{cloud}}$ is the total mass of the cloud (see Paper I or Table 8). $\sigma_{\mathrm{3d}}=\sqrt{3}\Delta V/2\sqrt{2\ln 2}$, where $\Delta V$ is the line width of the mean \xco spectrum of each sub-region (see Table 8).

\begin{deluxetable}{lrrrrrrrrrr}
\centering
\setlength\tabcolsep{3pt}
\tablecolumns{11}
\tabletypesize{\scriptsize}
\tablewidth{0pt}
\tablecaption{Global Physical Properties of the Sub-regions}
\tablehead{
 \colhead{Sub-region} & \colhead{$M_{\mathrm{cloud}}$\tablenotemark{a}} & \colhead{$R_{\mathrm{cloud}}$\tablenotemark{b}} & \colhead{$\Delta V$\tablenotemark{c}} & \colhead{$T$\tablenotemark{d}} & \colhead{$E_{\mathrm{turb}}$\tablenotemark{e}} & \colhead{$t_{\mathrm{ff}}$\tablenotemark{f}} & \colhead{$t_{\mathrm{diss}}$\tablenotemark{f}} & \colhead{$L_{\mathrm{turb}}$\tablenotemark{f}} & \colhead{$v_{\mathrm{esc}}$\tablenotemark{g}} & \colhead{$E_{\mathrm{grav}}$\tablenotemark{g}} \\
 \colhead{} & \colhead{($10^3$ M$_{\odot}$)} & \colhead{(pc)} & \colhead{(km s$^{-1}$)} & \colhead{(K)} & \colhead{(erg)} & \colhead{($10^6$ yr)} & \colhead{($10^6$ yr)} & \colhead{(erg s$^{-1}$)} & \colhead{(km s$^{-1}$)} & \colhead{(erg)}
}
\startdata
GGMC 1       &  36  & 32  & 2.4 &  8.3  & 1.1E+48 & 15.8 & 6.9  & 5.1E+34 & 3.1 & 3.5E+48 \\
GGMC 2       &  60  & 21  & 3.5 &  10.2 & 4.0E+48 &  6.5 & 4.9  & 2.6E+35 & 5.0 & 1.5E+49 \\
BFS 52       &  14  & 21  & 3.5 &  7.8  & 9.2E+47 & 13.5 & 4.2  & 7.1E+34 & 2.4 & 8.0E+47 \\
GGMC 3       &  110 & 27  & 4.0 &  10.2 & 9.5E+48 &  7.0 & 4.2  & 7.1E+35 & 5.9 & 3.8E+49 \\
GGMC 4       &  55  & 21  & 4.2 &  8.8  & 5.2E+48 &  6.8 & 3.5  & 4.8E+35 & 4.7 & 1.2E+49 \\
Local Lynds &  1   & 4   & 1.9 &  9.2  & 1.9E+46 &  4.2 & 1.8  & 3.4E+33 & 1.5 & 2.1E+46 \\
West Front  &  8   & 22  & 6.4 &  6.9  & 1.8E+48 & 19.1 & 0.8  & 6.5E+35 & 1.8 & 2.5E+47 \\
\enddata
\tablenotetext{a}{Cloud mass come from Paper I.}
\tablenotetext{b}{Cloud radius is estimated by the geometric mean of minor and major axes of extent of \xco integrated intensity emission.}
\tablenotetext{c}{\hspace{0.5pt}Line width of mean \xco spectra in the cloud.}
\tablenotetext{d}{Average excitation temperature of the cloud.}
\tablenotetext{e}{See Section 5.2.1.}
\tablenotetext{f}{See Section 5.2.2.}
\tablenotetext{g}{See Section 5.2.3.}
\end{deluxetable}

\begin{deluxetable}{lrrrrrr}
\centering
\setlength\tabcolsep{2pt}
\tablecolumns{7}
\tabletypesize{\scriptsize}
\tablewidth{0pt}
\tablecaption{The Impact of outflow candidates (Corrected) on its Parent Cloud}
\tablehead{
 \colhead{Cloud} & \colhead{$E_{\mathrm{flow}}/E_{\mathrm{turb}}$} & \colhead{$L_{\mathrm{flow}}/L_{\mathrm{turb}}$} & \colhead{$E_{\mathrm{flow}}/E_{\mathrm{grav}}$} & \colhead{$M_{\mathrm{esc}}/$M$_{\odot}$} & \colhead{$M_{\mathrm{esc}}/M_{\mathrm{flow}}$} & \colhead{$M_{\mathrm{esc}}/M_{\mathrm{cloud}}$}
}
\startdata
GGMC 1       & 1.2E-02 & 2.4E-02 & 4.0E-03 & 99  & 1.5 & 2.8E-03 \\
GGMC 2       & 1.2E-02 & 2.8E-02 & 3.1E-03 & 137 & 1.3 & 2.3E-03 \\
BFS 52       & 6.4E-03 & 8.5E-03 & 7.4E-03 & 60  & 1.8 & 4.3E-03 \\
GGMC 3       & 1.2E-02 & 2.2E-02 & 2.9E-03 & 236 & 1.3 & 2.1E-03 \\
GGMC 4       & 9.1E-03 & 2.4E-02 & 3.8E-03 & 99  & 2.0 & 1.8E-03 \\
Local Lynds & 3.2E-03 & 1.2E-02 & 2.9E-03 & 1   & 3.5 & 7.8E-04 \\
West Front  & 1.0E-03 & 1.0E-03 & 7.4E-03 & 16  & 3.3 & 2.0E-03 \\
\enddata
\end{deluxetable}

Comparing Table 6 with Table 8 regions by regions, the total kinetic energy of these detected outflow candidates in each sub-region is much less than the total turbulent energy of the corresponding cloud, and ratio of the two parameters in each sub-region is $\leq$ 1.2\% (see Table 9). The maximum $E_{\mathrm{flow}}/E_{\mathrm{turb}}=$ 19\% when we considered the SUFs. This indicates that outflows cannot provide enough energy to balance the turbulent energy in each sub-region at the scale of the whole cloud. Therefore, there must be some other sources of energy that support the turbulence, such as expanding shells, stellar winds, and gravity-driven turbulence \citep{ABG2011, KB2016}.

\citet{BHM2009} showed that outflows can be important on small scales using principal component analysis on CO line maps. They found that the scale of the regions influenced by small scale driving by outflows was $\sim$ 0.3--0.6 pc in NGC 1333, assuming a distance of 235 pc \citep{HBC2008}. Combined the \co and \xco (1-0) data from the FCRAO (single dish) and CARMA (interferometry) observations, \citet{PAC2013} suggested that outflows are likely important agents to maintain the turbulence in this region at a scale of $\sim$ 0.5 pc. Specifically, $E_{\mathrm{flow}}/E_{\mathrm{turb}}\sim$ 35\%, $L_{\mathrm{flow}}/L_{\mathrm{turb}}\sim 20$, $E_{\mathrm{flow}}/E_{\mathrm{grav}}\sim$ 1.4, where the definitions of the last two terms see Sections 5.2.2 and 5.2.3, respectively. Other studies, such as \citet{FL2002}, \citet{AS2006} and \citet{ABG2010}, have also shown that outflows have significant impact on their environments within few parsecs. The cloud scale in current study (several pc to dozens of pc) is too large. That could be the main reason why outflows are insufficient to provide required power to maintain turbulence.

\subsubsection{Impact of Outflow Candidates---Luminosity}

To estimate the impact of outflow candidates on the surrounding environment, another important way is to compare the total energy injection rate of outflow candidates (the total luminosity of outflow candidates) with the turbulence luminosity (the turbulent dissipation rate $L_{\mathrm{turb}}$) of the clouds. As mentioned above, the luminosity of a single candidate outflow lobe is $L_{\mathrm{lobe}}=E_{\mathrm{lobe}}/t_{\mathrm{lobe}}$, and the total luminosity in each sub-region is the sum of all lobe candidates located within the corresponding sub-region.

The turbulent dissipation rate $L_{\mathrm{turb}}$ is given by
\begin{equation}
 L_{\mathrm{turb}}=\frac{E_{\mathrm{turb}}}{t_{\mathrm{diss}}},
\end{equation}
where $t_{\mathrm{diss}}$ is the turbulent dissipation timescale. The numerical study of \citet{M1999} suggested that $t_{\mathrm{diss}}$ of the energy dissipation of uniformly driven magnetohydrodynamic turbulence is approximately given by:
\begin{equation}
 t_{\mathrm{diss}}=\left(\frac{3.9\kappa}{M_{\mathrm{rms}}}\right)t_{\mathrm{ff}},
\end{equation}
where $\kappa$ is the ratio of the driving wavelength over the Jean's length of the cloud $\lambda_{\mathrm{J}}$. $t_{\mathrm{ff}}$ is the free-fall timescale, and $M_{\mathrm{rms}}$ is the Mach number in the turbulence.

In simulations the turbulence driving length of a continuous outflow is approximately equal to the outflow lobe \citep[see][]{NL2007, CFC2009}, therefore the driving wavelength is set to be the mean outflow candidate length $\lambda_{\mathrm{d}}$, and
\begin{equation}
 \kappa=\lambda_{\mathrm{d}}/\lambda_{\mathrm{J}}.
\end{equation}
The Jean's length of the cloud is defined as
\begin{equation}
 \lambda_{\mathrm{J}}=c_{\mathrm{s}}\sqrt{\pi/G\rho_{\mathrm{reg}}},
\end{equation}
where $G$ is the gravitational constant, $c_{\mathrm{s}}$ is the sound speed and $\rho_{\mathrm{reg}}$ is the mass density of the specific cloud. For an ideal gas, $c_{\mathrm{s}}$ reads \citep{M1999}
\begin{equation}
  c_{\mathrm{s}}=(3kT/m_H\mu)^{1/2},
\end{equation}
where $T$ is the average excitation temperature of a specific cloud (see Table 8), $k$ is the Boltzmann's constant, $m_H$ is the mass of atomic hydrogen, and $\mu=2.72$ is the mean molecular weight \citep{B2010}. $\rho_{\mathrm{reg}}$ reads
\begin{equation}
 \rho_{\mathrm{reg}}=3M_{\mathrm{cloud}}/\left(4\pi R^3_{\mathrm{cloud}}\right).
\end{equation}
$M_{\mathrm{cloud}}$ and $R_{\mathrm{cloud}}$ are the mass and radius of the specific cloud (see Table 8). Finally,
\begin{equation}
 t_{\mathrm{ff}}=\sqrt{3\pi/32G\rho_{\mathrm{reg}}},
\end{equation}
and
\begin{equation}
  M_{\mathrm{rms}}=\sigma_{\mathrm{3d}}/c_{\mathrm{s}}.
\end{equation}
$\sigma_{\mathrm{3d}}$ see Section 5.2.1.

Using the formulas from (2) to (9), we determined the turbulent dissipation timescale and turbulent dissipation rate  of the clouds (see Table 8). Table 9 shows that the outflow candidate total luminosity is much less than the turbulent dissipation rate of the clouds suggesting that the observed outflow activity is not enough to support the measured turbulence in each sub-region, even if we considered the SUFs (the maximum $L_{\mathrm{flow}}/L_{\mathrm{turb}}$ becomes 38\%). The above estimation is not flawless. The major uncertainty of $L_{\mathrm{flow}}/L_{\mathrm{turb}}$ result from the uncertainties in the outflow candidate length and dynamical timescale.

As discussed in Section 5.2.1, the main reason why outflow candidates were insufficient to provide the required power to maintain turbulence may be that the cloud scales in the study were too large.

\subsubsection{Disruption of the Parent Cloud by Outflow Candidates}

The study of \citet{AG2002} shows that outflows can have a disruptive impact on their parent clouds, which can be evaluated by comparing the outflow energy with cloud gravitational binding energy ($E_{\mathrm{grav}}$), where
\begin{equation}
 E_{\mathrm{grav}}=\frac{GM^2_{\mathrm{cloud}}}{R_{\mathrm{cloud}}}.
\end{equation}
The gravitational binding energy of each cloud is listed in Table 8. Table 9 shows that the outflow candidate's total kinetic energy in each cloud occupies just a small fraction of cloud's gravitational binding energy (ranges from 0.3\% to 0.7\%), which indicates that outflow candidates cannot provide enough power to disrupt its parent cloud. The maximum $E_{\mathrm{flow}}/E_{\mathrm{grav}}$ only rises to 6\% if we considered the SUFs.

Another method to judge the importance of the disruptive effect of outflow candidates on clouds is using the ``escape mass'' $M_{\mathrm{esc}}$ \citep[see details in ][]{ABG2010}, which is given by
\begin{equation}
 M_{\mathrm{esc}}=\frac{P_{\mathrm{flow}}}{v_{\mathrm{esc}}}=\frac{P_{\mathrm{flow}}}{\sqrt{2GM_{\mathrm{cloud}}/R_{\mathrm{cloud}}}}.
\end{equation}
From Table 9, $M_{\mathrm{esc}}/M_{\mathrm{flow}}$ ranges from 1.3 to 3.5, which implies that the current outflow candidates have strong impacts on the environment immediately surrounding them within a small distance. However, in terms of $M_{\mathrm{esc}}/M_{\mathrm{cloud}}$, the current outflow candidate momentum could potentially disperse only $\leq$ 0.4\% of the mass in their respective regions (see Table 9). The maximum $E_{\mathrm{esc}}/E_{\mathrm{cloud}}$ only rises to 5\% if we considered the SUFs. All of these suggest that the current outflow candidates may disperse some gas from their parent clouds, but they do not have enough strength to make a serious impact at the scale of the whole cloud.

$E_{\mathrm{flow}}/E_{\mathrm{grav}}$ is $\sim$ 33\%--65\% at $\sim$ 0.2 pc scale in L1228 \citep{TM1997}, where the uncertainty of this ratio depends on the power-law dependence of the density law \citep{MRW1988}. This ratio is $\sim$ 4\%--40\% at the $\sim$ 0.6--2.0 pc scale in sub-regions within the Perseus molecular cloud complex \citep{ABG2010}, and $\sim$ 0.3\% at the $\sim$ 13.8 pc scale in the Taurus molecular cloud \citep{LLQ2015}. Therefore, scale is important for evaluating outflow impacts.

The scale of the interaction between the outflow candidates and clouds is of the order of dozens of pc in this study, hence it may not be suitable to discuss the impact of outflow candidates on the cores with size scales about a few tenths of a parsec. The results reported here do not rule out the possibility that outflows can be an important mechanism of mass-loss in cores \citep[see e.g.,][]{FL2002, AS2005, AS2006}.	

\section{Summaries and Conclusions}

A large-scale survey of outflow features was conducted of the GOS (a total of 58.5 square degrees) using \co and \xco molecular line maps, which are a part of MWISP.

We developed a set of semi-automated IDL scripts based on longitude-latitude-velocity (p-p-v) space to search for and evaluate outflows over a large area rapidly and efficiently. A total of 198 outflow candidates were identified, including 63 bipolar outflow candidates. Of these, 134 were located in the Perseus arm, which includes the Gem OB1, and 96\% of them were newly detected. Twenty-one were located in the Local arm, which includes the Local Lynds dark clouds and West Front clouds, and the remaining outflow candidates were beyond the Perseus arm. Owing to the limited resolution, especially in regions with distances of $\gtrsim$ 3.8 kpc, many of the identified outflow candidates were probably multi-outflow candidates.

We analyzed the distribution of outflow candidates and discussed the impact of outflows on its surrounding gas. The limitation of this study could result in underestimating outflow impact by mis-identifying outflow candidates. The main summaries and conclusions are as follows:

\begin{enumerate}
 \item Outflow candidates in GGMC 1, BFS 52 and Swallow were mainly associated with \xco intensity peaks or located at ring-like or filamentary structures. However, outflow candidates were ubiquitously distributed in GGMC 2, GGMC 3, GGMC 4, Local Lynds, West Front, Horn and Remote. Strong outflow candidates were also mainly located at filamentary or ring-like structures, or close to \xco intensity peaks or HII regions, but a few of them deviated from \xco intensity peaks.
 \item We obtained the ratio of the total outflow candidate kinetic energy to the turbulent energy of each host cloud, and the ratio of the outflow candidate luminosity (kinetic energy injection rate) to the luminosity (turbulence dissipation rate). Both of them were $\lesssim$ 3\%, and rose to $\lesssim$ 38\% if the undetected outflows were taken into consideration. These results showed that the identified outflow candidates cannot provide sufficient energy to balance turbulence for each sub-region, even undetected outflows were considered.
 \item The two ratios---the total outflow candidate kinetic energy divided by cloud's gravitational binding energy, and the ``escape mass'' (the cloud's mass potentially dispersed by outflow candidates) divided by cloud's mass---were both $\lesssim$ 1\%, and rose to $\lesssim$ 6\% if the undetected outflows were taken into consideration. The current outflow candidates may disperse some gas from their parent clouds, but did not have sufficient power to cause major disruptions to the cloud at the scale of the whole cloud, even accounting for undetected outflows.
\end{enumerate}

\acknowledgments
We thank the anonymous referee for many useful suggestions and comments that greatly improved this paper. This work was supported by the National Natural Science Foundation of China (Grant Numbers: 11673066 and 11233007), and the Key Laboratory for Radio Astronomy. The Starlink software \citep{CBJ2014} is supported by the East Asian Observatory.

\clearpage
\appendix

\section{Outflow Identification Methodology}

We conducted an unbiased, blind outflow survey toward studied region. To ensure efficiency and precision of the outflow identification, we developed a set of semi-automated IDL scripts for outflow searching. The scripts consist of seven steps. The example shown in this section is of sub-region GGMC 2.

\subsection{The Distribution of \xco Peak Velocity}

The first step is to extract spectral data for each pixel from three-dimensional \xco data of each sub-region. For each pixel, if the signal was larger than 3$\times$RMS in the velocity range, the velocity of the peak emission (peak velocity) would be extracted and a two-dimensional distribution map of \xco peak velocity would be created (see e.g. Figure A. 1). Note that, if two or more extreme brightness temperatures were not be confused with each other, the corresponding number of peak velocity was extracted and corresponding velocity distribution diagram was created. However, when some signals with extreme brightness temperatures were blended, or the differences between the peak velocities was less than the minimum $1\sigma$ velocity dispersion (Gaussian fit) of each peak, we adopted the mean value of the peak velocities as the systemic velocity. All \co outflow candidates searched here had corresponding \xco emission.

\begin{figure}[!ht]
\centering
\figurenum{A. 1}
\includegraphics[width=0.7\textwidth]{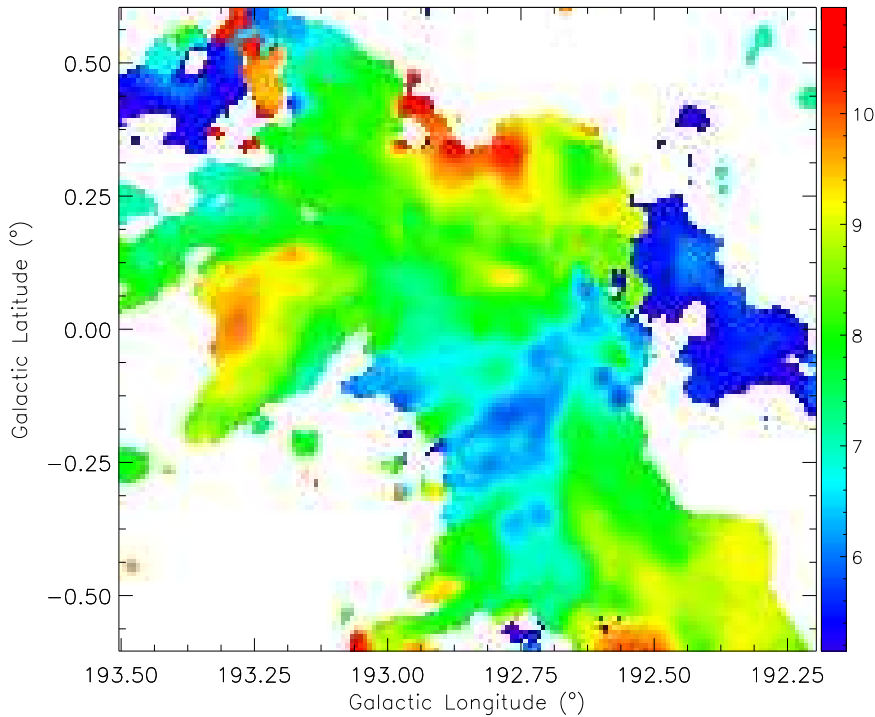}
\caption{\xco Peak-velocity in GGMC 2. The color-bar is in unit of km s$^{-1}$.}
\end{figure}

\subsection{The Distribution of Extensional Velocities of Blue and Red Line Wings}

Based on the first step, we then obtained the extensional velocities of the blue/red lobes of \co spectra of those pixels which had \xco emission and obtained \xco peak velocities. For a \co spectrum in a specific pixel, the scripts began with each \xco peak velocity, and searched the velocity forward or backward until \co emission in three successive channels were less than the threshold temperature, and this cut-off velocity was regarded as the extensional velocity of red/blue line wings.

To facilitate the subsequent analysis, we used the maximum velocity minus the extensional velocity of the blue lobe of each pixel in each sub-region to obtain a distribution of relative velocities of blue line wings, which were in positive values. Similarly, the relative velocities of the red line wings obtained by subtracting the minimum velocity from the extensional velocity of the red lobe of each pixel in each sub-region.

It is worthwhile to note that three threshold temperatures were contained to calculate the weighted mean extensional velocities of the blue/red lobes with the weighted portion of 3:3:4. The three threshold temperatures were 1$\times$RMS, 2$\times$RMS and 0.2$\times$peak temperature (if it was larger than 3$\times$RMS, we adopted 3$\times$RMS). The threshold temperature of 1$\times$RMS was chosen to avoid missing some faint outflows. We then created the distribution of the extensional velocities of the blue and red lobes. Figure A. 2 shows the distribution of extensional velocities of the red lobes in GGMC 2.

\subsection{The Positions of the Minimum Velocity Extents}

We used the CUPID \citep[][part of the STARLINK software$^8$]{BRJ2007} \footnotetext[8]{See \url{http://starlink.eao.hawaii.edu/starlink}.}clump-finding algorithm FELLWALKER \citep{B2015} to search for extreme points (i.e., the minimum velocity extents) from a distribution of extensional velocities of the blue and red lobes, and output their positions. The parameters setting in STARLINK CUPID FELLWALKER were based on the distance of a specific sub-region. The extensional velocity of a minimum velocity extent relative to ambient gases was set to be larger than 1 km s$^{-1}$. The preliminary positions of the blue and red lobes were obtained after we removed noise fluctuations. The positions of the red lobes in GGMC 2 are shown in Figure A. 2 as red pluses.

\begin{figure}[!t]
\centering
\figurenum{A. 2}
\includegraphics[width=0.7\textwidth]{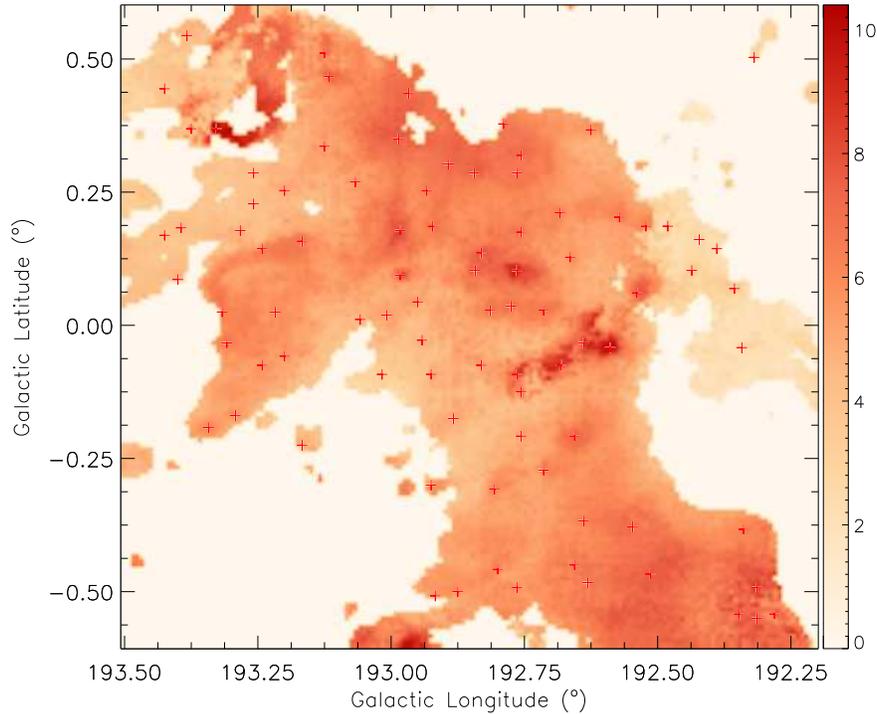}
\caption{The distribution of line wing maximum velocity in GGMC 2. Red pluses mark the preliminary positions of red lobes. The color-bar is in unit of km s$^{-1}$.}
\label{fig:sample_Uwingv}
\end{figure}

\subsection{Line Diagnoses of the Positions of the Minimum Velocity Extents}

Inevitably, the contaminations from stellar winds and weak clouds could influence the results. Therefore, we conducted line diagnostics at every position of the minimum velocity extents, and removed outflow candidates which were obviously incompatible with the line profiles of the outflow line wings.

We preliminary estimated the velocity range of line wings in specific spectrum. The maximum line wing velocity relative to the peak velocity was the extensional velocity of the \co line wing, and the minimum one was the extensional velocity of the \xco line wing or the maximum FWHM of $^{12}$CO. The line diagnostic criteria were as follows: first, to avoid mistaking core components as outflow components, the maximum velocity of a \co line wing relative to the peak velocity should be greater than 2 km s$^{-1}$ , and should be larger than the thermal velocity (taken as the FWHM of \co or the extensional velocity of the \xco line wing). Second, to remove the interfering components along the line-of-sight, the \co profile in the velocity range of the line wing should decrease successively.

In spite of such strict criteria promote the reliability of outflow candidate samples, it could miss the outflow candidates blended with other components. To guarantee the completeness of outflow candidate samples, we slightly loosened the second criterion that the \co profile could contain non-obvious bumps which covered less than 6 channels (i.e., 1.0 km s$^{-1}$). This step obtained outflow candidate samples.

\subsection{Plot Outflow Candidate's Contours and P-V Map Diagrams}

In this step, we plotted the integrated intensity map and contours over the entire line wing velocity range of the blue/red lobes for each outflow candidate, then plotted a position-velocity (P-V) diagram along four directions to investigate their velocity structure. This step was completed by eye. According to the contours and the P-V diagrams, we adjusted the line wing velocity ranges of the blue/red lobes or positions of outflow candidates to make the contours under an ideal state (i.e., make the score defined in A. 6 as high as possible). We then assigned the final position and the velocity range of outflow ling wings for each outflow candidate. Figure A. 3 and Figure A. 4 presents two examples from outflow candidates in GGMC 2 (red lobe of G2-3) and West Front (removed from our sample later), respectively. In Figure A. 3--A. 5, we arbitrarily chose $3\times3$ pixels window to average spectra (both for \co and $^{13}$CO) to improve the signal-to-noise ratio of the spectra. Maps like Figure A. 3 and Figure A. 4 are the direct sources of the scoring system for outflow candidates in the next subsection.

\begin{figure}[!t]
\centering
\figurenum{A. 3}
\includegraphics[width=0.6\textheight,angle=-90]{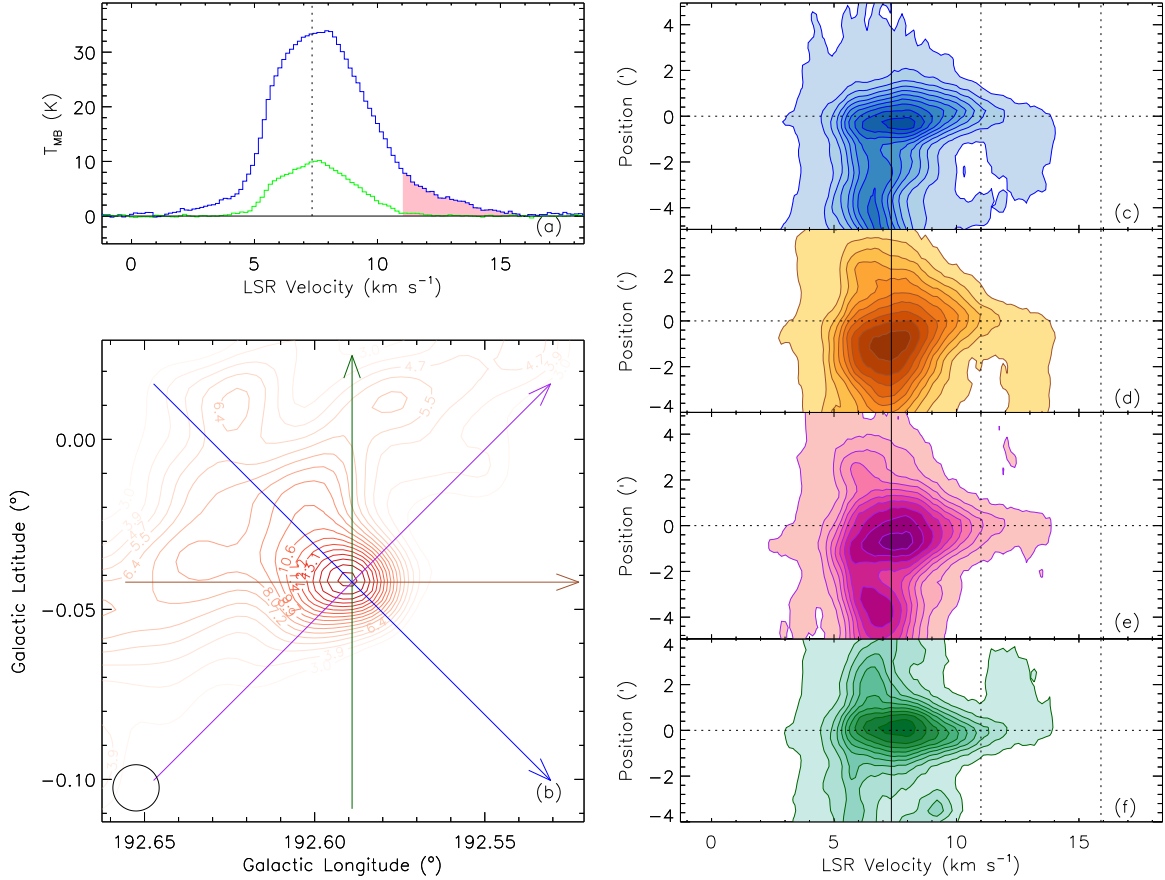}
\caption{Red lobe of G2-3 in GGMC 2. (a), the blue and green spectra are the \co and \xco averaged over $3\times3$ pixels centered at the emission peak of \co red lobe, respectively; and the pink shaded part of the spectrum indicates the \co line wing velocity range. (b), red \co lobe contours, the lowest contour of the lobes is 20\% of the peak intensity, the contour interval is 5\% of the peak intensity, and the color lightness of the contours increase with the intensity. (c)--(f), position-velocity (P-V) diagrams along the arrows with the same colors showed in panel (b), specifically, panels (c)--(f) correspond to the blue (point to southwest), yellow (point to west), purple (point to northwest) and green (point to north) arrows in panel (b), respectively; the width to draw the P-V diagram is 1$\arcmin$ (2 pixels); the lowest contour level of the P-V diagram is 5\% of the maximum value, as are the contour intervals. The open circle in the left-bottom corner indicates the size of HPBW of $^{12}$CO.}
\label{fig:sample_gem_out 1}
\end{figure}

\begin{figure}
\centering
\figurenum{A. 4}
\includegraphics[width=0.6\textheight,angle=-90]{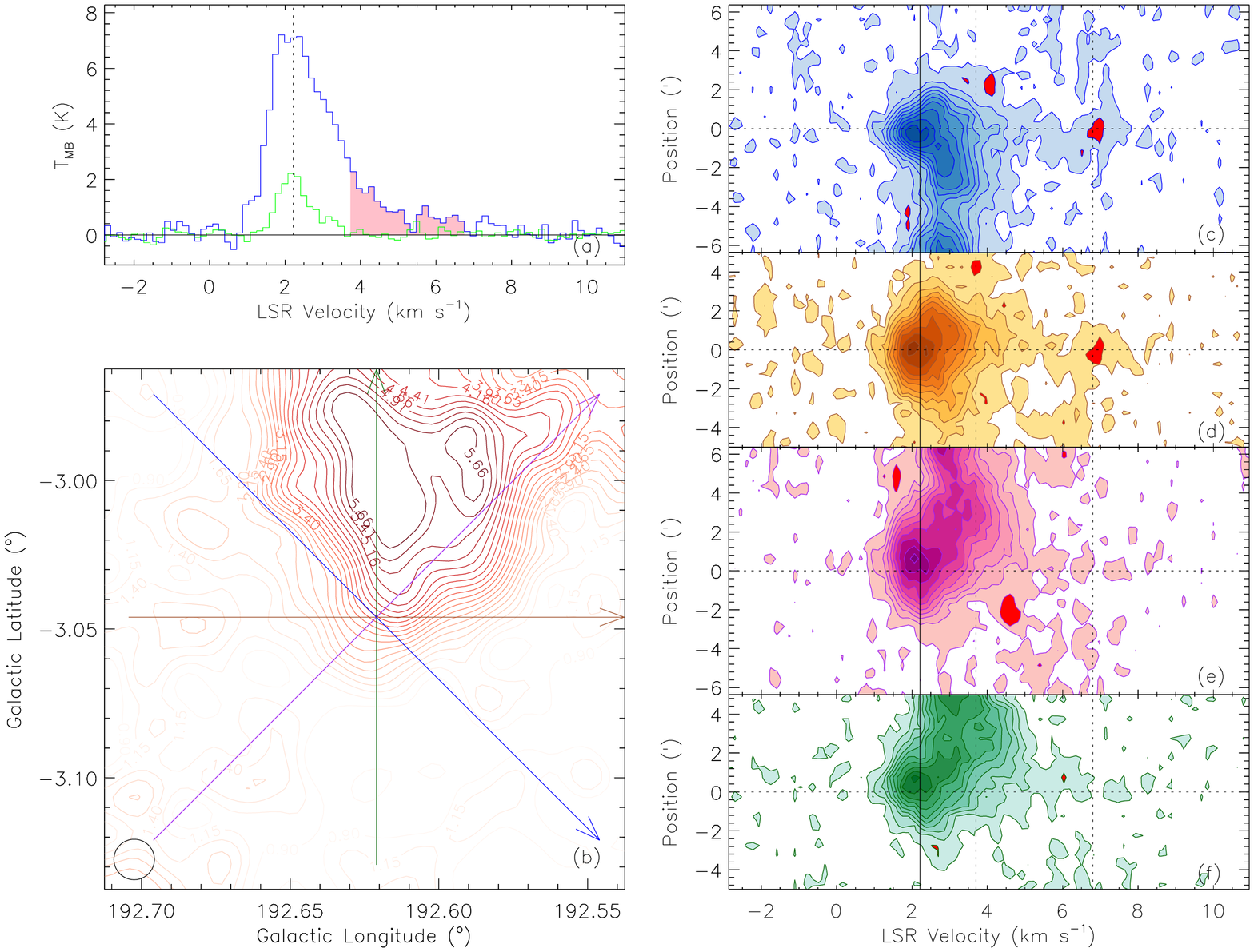}
\caption{A red lobe in West Front which was removed from our sample. The description of this figure refers to Figure A. 3.}
\label{fig:sample_gem_out 2}
\end{figure}

\subsection{Scoring System and bipolar outflow candidates}

To estimate the quality level of the outflow candidates obtained above, we set an artificial scoring system based on the line profiles, contour morphologies and P-V diagrams. Table A. 1 presents the detailed criteria. The overall score of a outflow lobe candidate calculated by adding all score of sub-items together. Table A. 2 presents the quality levels based on the overall scores.

Samples with quality levels of A are typically high-velocity outflows and the most reliable outflow candidates. Samples with quality levels of B are reliable, and the majority of outflow candidates belong to this classification. For samples with quality level C, the characteristics of the outflows are not obvious, but we cannot rule out the possibility of outflow activity in those candidates. We removed all outflow candidates with quality level D from our sample owing to their ambiguous outflow characteristics. The quality level is A for outflow candidate shown in Figure A. 3. However, for the example \citep[associated with CB 39 in][]{YC1992, WEZ1995} shown in Figure A. 4, the quality level is D, thus being removed from our sample.

\begin{deluxetable}{cccc}
\centering
\tablecolumns{4}
\tablenum{A. 1}
\tabletypesize{\small}
\tablewidth{0pt}
\setlength\tabcolsep{3pt}
\tablecaption{Detail Criterion of Scoring System}
\tablehead{
 \colhead{Line Profile} & \colhead{P-V map} & \colhead{Contours} & \colhead{Sub-Item Score} \\
 \colhead{Clear or Not}   & \colhead{Minimum Velocity Extent} & \colhead{Isolate or Not (Morphology)}  & \colhead{}
}
\startdata
Clear & Obvious Bulge & Absolute Isolate & 3 \\
Non-obvious bumps & Bulge with contamination & Differentiable & 2 \\
Contamination & No bulge or chaotic & Confused with Ambient Gas & 1
\enddata
\tablecomments{The four columns are the line profile of the line wings, P-V map, lobe contours, and sub-item score.}
\end{deluxetable}

\begin{deluxetable}{cc}
\centering
\tablecolumns{4}
\tablenum{A. 2}
\tabletypesize{\small}
\tablewidth{0pt}
\tablecaption{Quality Level}
\tablehead{
 \colhead{Composite Score} & \colhead{Quality Level}
}
\startdata
8-9 & A \\
6-7 & B \\
4-5 & C \\
3   & D
\enddata

\end{deluxetable}

In addition, blue and red lobes were paired up to form a bipolar outflow candidate if the two lobes were close (within 1.5 pc) or had similar core component structures (an example shows in Figure A. 5. of subsection A. 7). Column density of outflows (i.e., the number of outflows per square pc) was about 0.2 pc$^{-2}$ in previous studies \citep[e.g., ][]{ABG2010, LLQ2015}, therefore the average spatial interval of outflows was $\sim $ 4.5 pc indicating the threshold of 1.5 pc can be feasible.

Since the separation between the two lobes of R-6 is $\sim 5.1$ pc ($\sim 2\arcmin$), it is dubious to regard them as a bipolar outflow candidate. Another cases that remain to be determined include outflow candidates B-5 ($\sim 1.6$ pc apart), B-16 ($\sim 2.0$ pc apart), G3-8 ($\sim$ 2.0 pc apart), G3-11 ($\sim 1.6$ pc apart), G3-19 ($\sim 2.0$ pc apart), G3-27 ($\sim 2.0$ pc apart), G3-44 ($\sim$ 1.6 pc apart), G4-13 ($\sim 1.9$ pc apart), H-1 ($\sim 3.4$ pc apart), R-15 ($\sim 2.6$ pc apart) and R-18 ($\sim 2.1$ pc apart). We also cannot rule out the possibility that bipolar outflow candidates are several interacting outflows in some regions crowded with outflows (e.g., outflow candidates G3-21 and G3-22, etc.). Low spatial resolution of this work prevents further analysis of this problem. The total score and quality level of the bipolar outflow candidates were determined by the highest level.

\subsection{Outflow Exhibition and Comparison with WISE False Color Maps}

Figure A. 5 shows an example of outflow candidate G2-4 in GGMC 2. The green-filled contours in the left-bottom present the integrated intensity map of \xco core components superimposed by the contours of outflow lobes (blue/red for blue/red lobe). For bipolar outflow candidates, the integral velocity range from the incipient velocity of the blue line wing (starting with the peak velocity to the left) to the incipient velocity of the red line wing (starting with peak velocity to the right). For a monopolar outflow candidate, the integral velocity range is from the incipient velocity of blue/red line wing to its symmetrical point, and the symmetry axis is the peak velocity. The number of contour levels of \xco integrated intensity is 15, with a contour interval of 1/15 of the difference between the maximum and minimum integrated intensity in the mapped region. However, to sketch the contours of the main outflow candidate's structure, the incipient value of an outflow candidate's contour depends on the specific source, typical in the level of $\gtrsim 40$\% of the maximum integrated intensity in the mapped region.

\begin{figure}[!ht]
\figurenum{A. 5}
\centering
\includegraphics[height=0.48\textheight,angle=-90]{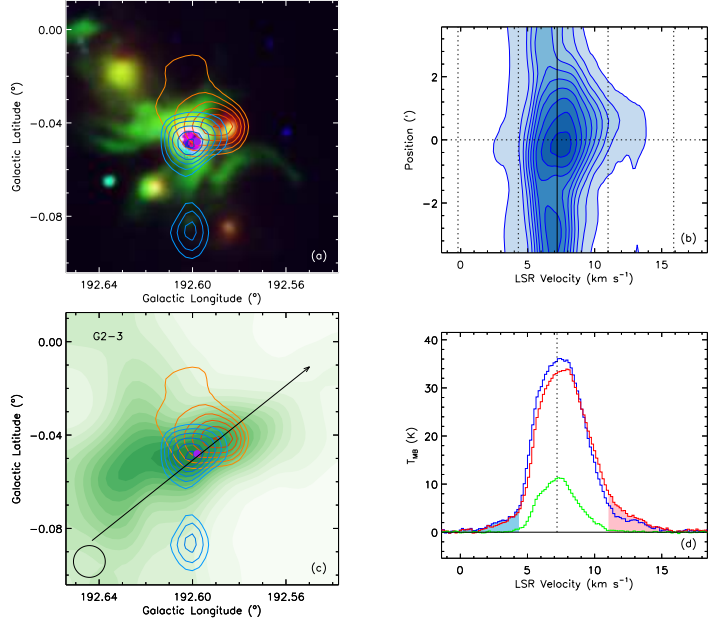}
\caption{An exhibition of outflow candidate G2-3 in GGMC 2. (a): Wise 4.6 (blue), 12 (green), 22 (red) $\mu$m false color image, and blue/red lobe contours. (b): a position-velocity (P-V) diagram along the black arrow shown in panel (c); the width to draw the P-V diagram is 1$\arcmin$ (2 pixels). (c): blue/red \co lobe contours overlaid on the integrated intensity of \xco cores depicted as green-filled contours. The black arrow begins from the position of a blue lobe and points to the red lobe for a bipolar outflow candidate, and for a monopolar outflow candidate the angle is 45$\degr$. (d): the blue/red spectrum is the \co averaged over $3\times3$ pixels centered at the blue/red emission peak positions, respectively; the green spectrum is the \xco averaged over $3\times3$ pixels centered at the purple point in the left-upper or left-bottom panel, which denotes the position of outflow candidate. The blue/red shaded part of the spectrum indicates the blue/red line wing velocity range of $^{12}$CO. The black dotted line in the center is of the peak velocity of $^{13}$CO. The open circle in the left-bottom corner shows the size of HPBW of $^{12}$CO.}
\label{fig:gem_G2-4}
\end{figure}

Infrared source is located at/near the center of outflows \citep{ZHB2001}. The high-sensitivity mid-infrared images from the Wide-field Infrared Survey Explorer \citep[WISE,][]{WEM2010} indicate the protostar activity that likely drive outflows. We therefore superimposed the outflow candidate contours on the WISE image and checked by eye for a point of light in the WISE image, see an example in Figure A. 5 (a). We only considered the \xco emission which associated with outflow lobe candidate (defined as \xco emission whose peak is the closest to the outflow lobe candidate). If there are WISE spots within the first five contour levels of \xco emission starting from the \xco peak emission, the outflow candidate was regarded as having WISE associations (denote as ``Y'' in the column ``WISE Detection'' in Table C). If there are WISE spots within $30\arcsec$ (the pixel size of the current study) of the lowest contour of the \co outflow lobe, the outflow candidate was regarded as likely having WISE associations (denote as ``Y?'' in the column ``WISE Detection'' in Table C). Figure 3 in Paper I indicated that there is little overlap among these 10 sub-regions in terms of \xco emission. Therefore, it is probable that the outflow candidates are associated with WISE emission (rather than happen to corresponding WISE spot lies along the line of sight). However, it is hard to rule out the possibility that the outflow candidates actually happen to coincide with a WISE spot in complicated cluster environments \citep[e.g., ][]{CZZ2016}, future high resolution investigation will be of great interest.

The outflow candidates reported here are likely to have WISE associations. In most cases, \co wing emission of outflow candidates is spatially confined (i.e., has limited spatial extent, e.g., the blue/red \co lobe contours in Figure A. 5), which is an essential feature of a typical outflow \citep{ZHB2001}.

\section{Calculation of Outflow Candidate's Parameters}

The areas and lengths of outflow lobe candidates ($A_{\mathrm{lobe}}$ and $l_{\mathrm{lobe}}$) can be estimated by their contours. As Figure A. 5 shows, generally speaking, the area of a relatively isolated lobe candidate is decided by the 40\% of peak integrated intensity. For outflow candidates which are contaminated by other components but can be resolved, we estimated the maximum areas of the vicinity of lobes where the outflow candidate components can be distinguished from other components. As a result of the limited resolution and great distance, the detailed structures of outflow candidates are usually unresolvable, so we calculated $l_{\mathrm{lobe}}$ by assuming a typical collimation factor of 2.45 \citep{WWZ2004}. The outflow lobe candidate length reads $l_{\mathrm{lobe}}=1.77\sqrt{A_{\mathrm{lobe}}}$. For a bipolar outflow candidate, its area and length were estimated by averaging the area and length of its two lobes.

Similar to \citet{BSS2002}, we adopted the line wing velocity which was the furthest away from the central systemic velocity to compute the dynamical timescale of outflow lobe candidates, see as follow:
\begin{equation}
 t_{\mathrm{lobe}}=\frac{l_{\mathrm{lobe}}}{\Delta v_{\mathrm{max}}},
\end{equation}
where $l_{\mathrm{lobe}}$ is the outflow lobe candidate length, and $\Delta v_{\mathrm{max}}$ is the maximum outflow lobe candidate velocity. The value of $\Delta v_{\mathrm{max}}$ presented here is based on the sensitivity of our data, i.e., 0.45 K.

To estimate the mass of each lobe candidate, we computed the column density of each \co lobe line wing. Referring to \citet{GRB1990}, and assuming that the excitation of all energy levels are the same, the column density of a linear, rigid rotor molecular from one transition under conditions of LTE is
\begin{equation}
 N=\frac{3k}{8\pi^3B\mu^2}\frac{\exp[hBJ(J+1)/kT_{\mathrm{ex}}]}{J+1}\frac{T_{\mathrm{ex}}+hB/3k}{1-\exp({-h\nu/kT_{\mathrm{ex}}})}\int{\tau_{\nu}}dv,
\end{equation}
where $J$ is the quantum number of the lower rotational level, and $B$ and $\mu$ are the rotational constant and permanent dipole moment, respectively. For $^{12}$CO, we assume that the high-velocity gas is optically thin, and the area filling factor $f=1$, the total column density reads \citep[see][]{SSS1984}
\begin{equation}
 N(^{12}\mathrm{CO})=4.2\times 10^{13}\frac{T_{\mathrm{ex}}}{\exp(-5.5/T_{\mathrm{ex}})}\int{T_{\mathrm{mb}}}dv,
\end{equation}
where the integrated range is the wing range. The excitation temperature $T_{\mathrm{ex}}$ is assumed to be 30 K. \citet{AG2001}, \citet{ABG2010} used the main beam temperature in core centers to compute $T_{\mathrm{ex}}$; however, this approach would underestimate $T_{\mathrm{ex}}$ for the beam dilution effect in our moderate resolution.

Although we have \co and \xco data, we did not estimate the optical depth, because in most cases \xco emission is not strong enough as a signal in the velocity range of the \co line wing. Based on the assumption of optical thin \co line wing, the column density is underestimated to some extent. However, in fact, the errors of the column density arising from other factors are more significant, including: the error of the interception of outflow candidate velocity (for example, ignoring the low-velocity outflow candidate component that is blended with cores), and the error of the estimation of lobe candidate area.

The column density of H$_2$ gas, $N$, can be calculated by multiplying $N(^{12}\mathrm{CO})$ by a conversion factor $[\mathrm{H}]/[^{12}\mathrm{CO}]$ $=1\times 10^4$ \citep[see e.g., ][]{SSS1984}. The mass of the outflow lobe candidate is as follow:
\begin{equation}
 M_{\mathrm{lobe}}=(N_{\mathrm{lobe}}\times A_{\mathrm{lobe}})m_{\mathrm{H_2}},
\end{equation}
where $M_{\mathrm{lobe}}$, $A_{\mathrm{lobe}}$ and $N_{\mathrm{lobe}}$ are the mass, area and column density of blue/red lobe of outflow candidates, respectively, and $m_{\mathrm{H_2}}$ is the mass of a hydrogen molecule. If the outflow candidate is bipolar, the mass $M_{\mathrm{out}}$ is the sum of the masses of blue and red lobes. However, in order to avoid inconsistent in statistics, we chose the maximum mass of the two lobes for a bipolar outflow candidate when we compared outflow candidates to each other. Note that the estimated mass is a lower limit for the outflow candidate.

The estimation of the momentum and energy of an outflow candidate are based on the outflow candidate velocity relative to the central cloud. We calculated the lobe candidate velocity in each spatial pixel using the temperature-weighted mean velocity of the wing channels, as follow:
\begin{equation}
 \langle\Delta v_{\mathrm{lobe}}\rangle=\frac{\displaystyle\sum_{i}\left(v_i-v_{\mathrm{peak}}\right)T_i\Delta v_{\mathrm{res}}}{\displaystyle\sum_{i}T_i\Delta v_{\mathrm{res}}},
\end{equation}
where $i$ and $T_i$ are the channel index of the blue/red wing and the brightness temperature of each channel, and $\Delta v_{\mathrm{res}}$ is the velocity resolution of a channel. The square of lobe candidate velocity can be obtained similarly through
\begin{equation}
 \langle\Delta v^2_{\mathrm{lobe}}\rangle=\frac{\displaystyle\sum_{i}\left(v_i-v_{\mathrm{peak}}\right)^2T_i\Delta v_{\mathrm{res}}}{\displaystyle\sum_{i}T_i\Delta v_{\mathrm{res}}},
\end{equation}
And the momentum, energy and luminosity of an outflow lobe candidate are
\begin{equation}
 P_{\mathrm{lobe}}=\displaystyle\sum_{A_{\mathrm{lobe}}}M_{\mathrm{lobe}}\langle\Delta v_{\mathrm{lobe}}\rangle,
\end{equation}
\begin{equation}
 E_{\mathrm{lobe}}=\frac{1}{2}\displaystyle\sum_{A_{\mathrm{lobe}}}M_{\mathrm{lobe}}\langle\Delta v^2_{\mathrm{lobe}}\rangle,
\end{equation}
\begin{equation}
 L_{\mathrm{lobe}}=E_{\mathrm{lobe}}/t_{\mathrm{lobe}}.
\end{equation}
For a bipolar outflow candidate, the velocity was the mean value of the two lobes, but the momentum, energy and luminosity were the sum values. However, in order to avoid inconsistent in statistics, we chose the maximum velocity, momentum, energy and luminosity of the two lobes for a bipolar outflow candidate when we compared outflow candidates to each other.

\clearpage
\section{Outflow Samples}

This appendix presents the outflows with possible WISE associations (total of 198) towards GOS. The quality level A denotes very reliable outflows, B denotes reliable outflows, and C denotes outflows whose characteristics are less obvious. In the column ``Wise Detection'', ``Y'' present outflows that are associated with a WISE source, ``Y?'' present outflows are possible associated with a WISE source.

\begin{deluxetable}{lrrccccc}
\centering
\setlength\tabcolsep{3pt}
\tablecolumns{8}
\tabletypesize{\small}
\tablewidth{0pt}
\tablenum{C}
\tablecaption{Outflow Sample Statistics}
\tablehead{
\colhead{Index} & \colhead{$l$} &  \colhead{$b$} & \colhead{Blue Line Wing}  & \colhead{Red Line Wing} & \colhead{Quality Level} & \colhead{WISE Detection} & \colhead{New Detection}  \\
 \colhead{}  & \colhead{($\degr$)} &  \colhead{($\degr$)} & \colhead{(km s$^{-1}$)} & \colhead{(km s$^{-1}$)} & \colhead{Blue$|$Red} & \colhead{} & \colhead{}
}
\startdata
\multicolumn{8}{l}{GGMC 1} \\
\hline
G1-1  & 192.720 & -1.245 & \nodata          & $(    5.9, 7.2)$ & C     &  Y?  &  Y \\
G1-2  & 192.842 & -1.133 & \nodata          & $(    5.8, 7.0)$ & B     &  Y?  &  Y \\
G1-3  & 193.093 & -1.388 & \nodata          & $(    5.5, 8.0)$ & C     &  Y?  &  Y \\
G1-4  & 193.238 & -1.619 & $(    1.5, 3.0)$ & $(    5.1, 6.5)$ & C$|$C &  Y?  &  Y \\
G1-5  & 193.250 & -1.038 & \nodata          & $(    5.2, 6.3)$ & C     &  Y?  &  Y \\
G1-6  & 193.310 & -1.374 & $(   -1.1, 1.0)$ & $(    4.5, 7.0)$ & C$|$C &  Y   &  Y \\
G1-7  & 193.497 & -1.155 & $(   -0.5, 1.3)$ & $(    4.0, 6.2)$ & C$|$B &  Y?  &  Y \\
G1-8  & 193.506 & -0.029 & \nodata          & $(    5.6, 6.8)$ & B     &  Y?  &  Y \\
G1-9  & 193.617 & -0.138 & $(    0.8, 2.0)$ & \nodata          & C     &  Y?  &  Y \\
G1-10 & 193.650 & -0.610 & $(   -0.6, 1.0)$ & \nodata          & C     &  Y?  &  Y \\
G1-11 & 193.680 &  0.185 & \nodata          & $(    8.5, 9.5)$ & B     &  Y?   &  Y \\
G1-12 & 193.717 & -0.275 & $(    0.0, 2.5)$ & \nodata          & C     &  Y   &  Y \\
G1-13 & 193.767 & -0.800 & $(    0.0, 1.3)$ & \nodata          & C     &  Y   &  Y \\
G1-14 & 193.845 & -0.395 & $(    1.1, 3.0)$ & \nodata          & B     &  Y?  &  Y \\
G1-15 & 193.885 & -1.086 & $(   -1.9, 1.5)$ & \nodata          & C     &  Y?  &  Y \\
G1-16 & 193.908 &  0.205 & $(    1.9, 3.4)$ & \nodata          & C     &  Y   &  Y \\
G1-17 & 193.915 & -1.000 & $(    0.0, 1.8)$ & \nodata          & B     &  Y   &  Y \\
G1-18 & 193.998 &  0.238 & $(    1.4, 2.7)$ & \nodata          & B     &  Y?  &  Y \\
G1-19 & 194.017 & -0.729 & $(   -0.6, 1.1)$ & \nodata          & C     &  Y?  &  Y \\
G1-20 & 194.020 & -1.095 & \nodata          & $(    5.8, 7.8)$ & A     &  Y?  &  Y \\
G1-21 & 194.025 & -0.875 & \nodata          & $(    5.0, 7.5)$ & B     &  Y?  &  Y \\
G1-22 & 194.056 & -0.663 & $(   -1.1, 1.3)$ & $(    5.0, 6.5)$ & B$|$C &  Y?  &  Y \\
G1-23 & 194.062 & -0.238 & $(    0.0, 2.0)$ & \nodata          & B     &  Y?  &  Y \\
G1-24 & 194.110 & -0.485 & \nodata          & $(    5.1, 6.7)$ & A     &  Y   &  Y \\
G1-25 & 194.150 & -0.792 & \nodata          & $(    4.6, 7.0)$ & B     &  Y?  &  Y \\
G1-26 & 194.175 & -0.165 & $(    0.0, 1.8)$ & \nodata          & B     &  Y?  &  Y \\
G1-27 & 194.180 & -1.100 & \nodata          & $(    5.5, 7.5)$ & B     &  Y?  &  Y \\
G1-28 & 194.185 & -0.935 & \nodata          & $(    5.0, 7.1)$ & A     &  Y?  &  Y \\
G1-29 & 194.235 &  0.113 & $(   -1.1, 1.0)$ & \nodata          & B     &  Y?  &  Y \\
G1-30 & 194.246 & -0.196 & $(    0.5, 1.8)$ & \nodata          & B     &  Y?  &  Y \\
G1-31 & 194.310 & -0.263 & $(    0.0, 2.3)$ & \nodata          & C     &  Y   &  Y \\
G1-32 & 194.335 & -0.990 & $(   -0.5, 1.0)$ & \nodata          & B     &  Y?  &  Y \\
G1-33 & 194.370 & -0.790 & $(    0.5, 2.0)$ & $(    4.5, 5.5)$ & B$|$C &  Y?  &  Y \\
G1-34 & 194.661 &  0.319 & $(    3.5, 5.8)$ & $(    9.0,11.0)$ & B$|$B &  Y   &  Y \\
\hline
\multicolumn{8}{l}{GGMC 2} \\
\hline
G2-1  & 192.455 &  0.005 & $(    1.2, 3.5)$ & \nodata          & B     &  Y?  &  Y \\
G2-2  & 192.580 &  0.210 & \nodata          & $(   10.2,12.4)$ & B     &  Y?  &  Y \\
G2-3  & 192.598 & -0.048 & $(   -0.2, 4.3)$ & $(   11.0,15.9)$ & A$|$A &  Y   &  N\tablenotemark{a} \\
G2-4  & 192.608 & -0.129 & $(    0.5, 4.3)$ & \nodata          & C     &  Y?  &  Y \\
G2-5  & 192.655 & -0.210 & \nodata          & $(   11.0,13.0)$ & A     &  Y?  &  Y \\
G2-6  & 192.695 &  0.201 & $(    6.0, 7.2)$ & $(   10.0,11.4)$ & B$|$B &  Y   &  Y \\
G2-7  & 192.715 & -0.254 & $(    1.1, 4.0)$ & $(    9.5,11.6)$ & B$|$C &  Y   &  Y \\
G2-8  & 192.718 &  0.025 & \nodata          & $(    9.5,12.2)$ & A     &  Y?  &  Y \\
G2-9  & 192.730 & -0.070 & $(    2.2, 4.5)$ & \nodata          & B     &  Y?  &  Y \\
G2-10 & 192.771 & -0.197 & $(    1.3, 4.0)$ & $(    9.4,12.2)$ & B$|$B &  Y?  &  Y \\
G2-11 & 192.820 &  0.125 & \nodata          & $(   11.2,13.8)$ & C     &  Y   &  Y \\
G2-12 & 192.846 &  0.288 & \nodata          & $(   11.3,12.9)$ & B     &  Y?  &  Y \\
G2-13 & 192.875 & -0.060 & \nodata          & $(    8.6,10.8)$ & B     &  Y?  &  Y \\
G2-14 & 192.932 &  0.192 & $(    2.0, 5.0)$ & $(   10.0,13.0)$ & A$|$C &  Y   &  Y \\
G2-15 & 192.979 &  0.092 & \nodata          & $(    9.5,15.2)$ & A     &  Y   &  Y \\
G2-16 & 192.992 &  0.162 & $(   -1.0, 5.3)$ & $(   11.0,14.8)$ & B$|$A &  Y?  &  N\tablenotemark{a} \\
G2-17 & 193.090 &  0.208 & $(    4.1, 6.0)$ & \nodata          & C     &  Y?  &  Y \\
G2-18 & 193.197 &  0.254 & \nodata          & $(    9.0,10.6)$ & B     &  Y?  &  Y \\
G2-19 & 193.288 &  0.185 & $(    4.0, 6.0)$ & $(    8.8,10.4)$ & C$|$C &  Y?  &  Y \\
G2-20 & 193.320 &  0.030 & \nodata          & $(   10.8,12.0)$ & B     &  Y?  &  Y \\
\hline
\multicolumn{8}{l}{BFS 52} \\
\hline
B-1   & 191.066 &  0.819 & $(    2.4, 5.6)$ & $(    9.5,11.4)$ & C$|$A &  Y   &  Y \\
B-2   & 191.083 &  0.740 & \nodata          & $(    9.5,10.5)$ & B     &  Y?  &  Y \\
B-3   & 191.137 &  0.662 & \nodata          & $(    7.1, 9.5)$ & B     &  Y?  &  Y \\
B-4   & 191.220 &  0.752 & \nodata          & $(    7.0, 9.0)$ & B     &  Y?   &  Y \\
B-5   & 191.304 &  0.956 & $(    5.7, 6.8)$ & $(    9.5,10.8)$ & B$|$A &  Y?  &  Y \\
B-6   & 191.500 &  0.615 & \nodata          & $(   10.2,11.5)$ & B     &  Y   &  Y \\
B-7   & 191.742 &  0.833 & $(    4.3, 6.3)$ & \nodata          & B     &  Y?  &  Y \\
B-8   & 191.796 &  0.785 & $(    4.7, 5.7)$ & $(    8.0, 9.7)$ & B$|$C &  Y?  &  Y \\
B-9   & 191.827 &  1.000 & \nodata          & $(   10.2,11.7)$ & B     &  Y?  &  Y \\
B-10  & 191.895 &  0.778 & $(    4.9, 6.0)$ & \nodata          & C     &  Y?  &  Y \\
B-11  & 191.896 &  1.095 & $(    7.0, 8.5)$ & $(   11.2,12.2)$ & C$|$A &  Y?   &  Y \\
B-12  & 191.905 &  0.894 & $(    4.1, 6.0)$ & $(   10.0,13.6)$ & A$|$A &  Y   &  Y \\
B-13  & 192.030 &  1.046 & $(    5.2, 6.5)$ & $(    8.6,10.2)$ & A$|$A &  Y   &  Y \\
B-14  & 192.742 &  1.282 & $(    6.5, 7.6)$ & \nodata          & B     &  Y?  &  Y \\
B-15  & 193.009 &  1.067 & $(    7.6, 8.8)$ & \nodata          & C     &  Y?  &  Y \\
B-16  & 193.139 &  0.700 & $(    3.3, 5.0)$ & $(    9.6,10.8)$ & C$|$C &  Y?  &  Y \\
B-17  & 193.161 &  0.944 & $(    7.3, 8.5)$ & $(   12.0,13.5)$ & A$|$B &  Y?   &  Y \\
\hline
\multicolumn{8}{l}{GGMC 3} \\
\hline
G3-1  & 189.221 &  0.846 & \nodata          & $(   11.5,14.1)$ & A     &  Y   &  Y \\
G3-2  & 189.225 &  0.767 & \nodata          & $(   11.1,12.2)$ & B     &  Y   &  Y \\
G3-3  & 189.237 & -0.529 & $(    6.7, 8.0)$ & \nodata          & B     &  Y?  &  Y \\
G3-4  & 189.354 & -0.413 & $(    5.7, 8.2)$ & \nodata          & B     &  Y?  &  Y \\
G3-5  & 189.383 &  0.463 & \nodata          & $(   12.8,15.9)$ & A     &  Y?  &  Y \\
G3-6  & 189.580 & -0.145 & \nodata          & $(   10.5,12.1)$ & A     &  Y   &  Y \\
G3-7  & 189.583 &  0.000 & \nodata          & $(   11.5,15.4)$ & A     &  Y   &  Y \\
G3-8  & 189.609 &  0.170 & $(    1.5, 3.5)$ & $(   11.0,12.7)$ & B$|$C &  Y   &  Y \\
G3-9  & 189.633 &  0.085 & \nodata          & $(   11.0,13.0)$ & B     &  Y?  &  Y \\
G3-10 & 189.704 &  0.067 & $(    1.3, 4.0)$ & \nodata          & B     &  Y   &  Y \\
G3-11 & 189.729 &  0.023 & $(    1.6, 3.5)$ & $(   10.8,12.5)$ & C$|$B &  Y?  &  Y \\
G3-12 & 189.742 & -0.040 & \nodata          & $(    9.8,11.9)$ & B     &  Y?  &  Y \\
G3-13 & 189.747 &  0.165 & $(    1.0, 4.0)$ & \nodata          & B     &  Y?  &  Y \\
G3-14 & 189.790 &  0.344 & $(   -0.5, 3.8)$ & $(   11.8,15.9)$ & A$|$A &  Y   &  N\tablenotemark{b} \\
G3-15 & 189.792 &  0.467 & \nodata          & $(   11.3,13.8)$ & A     &  Y?  &  Y \\
G3-16 & 189.804 & -0.625 & \nodata          & $(    9.4,10.8)$ & B     &  Y?  &  Y \\
G3-17 & 189.821 &  0.088 & $(    0.3, 3.7)$ & \nodata          & A     &  Y?  &  Y \\
G3-18 & 189.833 &  0.313 & \nodata          & $(   11.8,15.7)$ & B     &  Y   &  Y \\
G3-19 & 189.842 &  0.194 & $(    1.4, 3.5)$ & $(   12.5,14.6)$ & B$|$A &  Y?  &  Y \\
G3-20 & 189.942 &  0.225 & $(    1.9, 3.8)$ & \nodata          & A     &  Y   &  Y \\
G3-21 & 189.973 &  0.331 & $(    0.8, 5.5)$ & $(   10.8,15.1)$ & B$|$A &  Y?  &  Y \\
G3-22 & 190.008 & -0.138 & \nodata          & $(   13.0,15.7)$ & B     &  Y?  &  Y \\
G3-23 & 190.017 &  0.125 & $(    2.5, 4.8)$ & \nodata          & B     &  Y?  &  Y \\
G3-24 & 190.030 & -0.665 & $(    2.4, 4.5)$ & \nodata          & A     &  Y?  &  Y \\
G3-25 & 190.045 &  0.522 & $(    3.5, 6.5)$ & $(   10.3,11.7)$ & A$|$A &  Y   &  Y \\
G3-26 & 190.081 &  0.246 & $(    1.1, 4.0)$ & $(   11.0,12.5)$ & A$|$A &  Y?  &  Y \\
G3-27 & 190.094 &  0.663 & $(    1.0, 4.6)$ & $(   10.1,12.2)$ & B$|$B &  Y?  &  Y \\
G3-28 & 190.101 &  0.604 & $(    3.5, 5.5)$ & $(    9.5,11.2)$ & A$|$A &  Y?   &  Y \\
G3-29 & 190.167 &  0.735 & $(    2.5, 4.6)$ & $(   10.3,12.2)$ & A$|$A &  Y   &  Y \\
G3-30 & 190.181 &  0.172 & $(    1.9, 4.8)$ & $(    9.5,11.0)$ & A$|$B &  Y?   &  Y \\
G3-31 & 190.185 &  0.365 & \nodata          & $(   10.0,11.5)$ & A     &  Y?   &  Y \\
G3-32 & 190.255 &  0.900 & $(    4.6, 7.6)$ & $(   10.7,12.9)$ & B$|$B &  Y?  &  Y \\
G3-33 & 190.275 &  0.365 & \nodata          & $(    9.3,11.3)$ & B     &  Y?  &  Y \\
G3-34 & 190.337 &  0.046 & $(    3.2, 5.6)$ & \nodata          & B     &  Y?  &  Y \\
G3-35 & 190.410 &  0.960 & $(    2.4, 4.6)$ & \nodata          & C     &  Y?  &  Y \\
G3-36 & 190.415 &  0.620 & $(    4.0, 6.0)$ & \nodata          & C     &  Y?   &  Y \\
G3-37 & 190.546 &  1.100 & $(    3.0, 4.5)$ & \nodata          & B     &  Y?  &  Y \\
G3-38 & 190.558 &  0.813 & $(    2.4, 6.5)$ & \nodata          & A     &  Y   &  Y \\
G3-39 & 190.562 &  0.600 & \nodata          & $(    9.0,10.6)$ & B     &  Y?  &  Y \\
G3-40 & 190.567 &  0.743 & \nodata          & $(   10.5,14.1)$ & A     &  Y   &  Y \\
G3-41 & 190.692 &  0.425 & \nodata          & $(    7.8, 9.8)$ & C     &  Y?  &  Y \\
G3-42 & 190.738 &  0.805 & $(    2.0, 3.5)$ & \nodata          & B     &  Y?  &  Y \\
G3-43 & 190.775 &  0.696 & \nodata          & $(   12.1,13.5)$ & C     &  Y?  &  Y \\
G3-44 & 190.918 &  0.321 & $(    2.8, 5.0)$ & $(    8.2, 9.3)$ & A$|$B &  Y?   &  Y \\
G3-45 & 191.052 &  0.361 & $(    4.6, 6.0)$ & $(    9.0,10.5)$ & B$|$B &  Y?   &  Y \\
\hline
\multicolumn{8}{l}{GGMC 4} \\
\hline
G4-1  & 188.673 &  0.935 & \nodata          & $(    4.8, 6.0)$ & B     &  Y?  &  Y \\
G4-2  & 188.790 &  1.290 & $(   -2.1,-0.5)$ & \nodata          & B     &  Y?   &  Y \\
G4-3  & 188.796 &  1.026 & $(   -5.0,-3.0)$ & \nodata          & A     &  Y   &  Y \\
G4-4  & 188.812 &  1.063 & $(   -5.9,-3.2)$ & \nodata          & A     &  Y   &  Y \\
G4-5  & 188.817 &  0.933 & \nodata          & $(    5.5, 6.8)$ & B     &  Y   &  Y \\
G4-6  & 188.940 &  1.060 & $(   -3.0,-1.0)$ & \nodata          & B     &  Y?  &  Y \\
G4-7  & 188.943 &  0.883 & $(   -6.0,-0.5)$ & $(    7.0,14.6)$ & A$|$A &  Y   &  N\tablenotemark{c} \\
G4-8  & 188.979 &  1.125 & $(   -5.7,-1.8)$ & \nodata          & B     &  Y?  &  Y \\
G4-9  & 189.000 &  1.213 & $(   -6.0,-2.8)$ & \nodata          & A     &  Y?  &  Y \\
G4-10 & 189.040 &  0.800 & $(   -7.0,-1.0)$ & $(    6.0,10.8)$ & A$|$A &  Y   &  N\tablenotemark{a} \\
G4-11 & 189.058 &  1.079 & $(   -4.4,-1.0)$ & \nodata          & A     &  Y   &  Y \\
G4-12 & 189.160 &  0.455 & $(   -0.0, 1.0)$ & \nodata          & B     &  Y?  &  Y \\
G4-13 & 189.201 &  0.373 & $(    0.2, 1.2)$ & $(    3.5, 4.9)$ & C$|$C &  Y?   &  Y \\
G4-14 & 189.268 &  1.285 & \nodata          & $(    8.5,10.2)$ & C     &  Y?  &  Y \\
G4-15 & 189.400 &  1.783 & $(   -2.5,-0.5)$ & \nodata          & B     &  Y?  &  Y \\
G4-16 & 189.454 &  0.738 & \nodata          & $(    6.5, 8.1)$ & C     &  Y?  &  Y \\
G4-17 & 189.575 &  0.968 & \nodata          & $(    6.2, 8.0)$ & C     &  Y?  &  Y \\
G4-18 & 189.687 &  1.779 & $(   -3.7,-1.7)$ & \nodata          & B     &  Y?  &  Y \\
\hline
\multicolumn{8}{l}{Local Lynds} \\
\hline
L-1   & 190.660 & -0.400 & $(   -3.7,-1.5)$ & \nodata          & B     &  Y?  &  Y \\
L-2   & 192.024 & -0.964 & $(   -6.2,-3.5)$ & $(    1.5, 3.0)$ & A$|$B &  Y?  &  Y \\
\hline
\multicolumn{8}{l}{West Front} \\
\hline
W-1   & 187.092 & -0.352 & $(   -2.9,-1.5)$ & \nodata          & C     &  Y   &  Y \\
W-2   & 187.535 & -0.728 & $(   -3.3,-0.3)$ & \nodata          & C     &  Y?  &  Y \\
W-3   & 187.562 & -3.211 & $(   -1.9, 0.3)$ & $(    3.7, 4.8)$ & B$|$B &  Y   &  Y \\
W-4   & 188.271 & -2.868 & $(    1.5, 2.5)$ & $(    4.2, 6.2)$ & B$|$B &  Y?  &  Y \\
W-5   & 188.970 & -1.940 & $(   -6.7,-3.5)$ & \nodata          & A     &  Y   &  Y \\
W-6   & 189.244 & -1.583 & $(   -1.7, 0.2)$ & \nodata          & C     &  Y?  &  Y \\
W-7   & 190.133 & -2.420 & \nodata          & $(    6.0, 8.3)$ & C     &  Y?   &  Y \\
W-8   & 190.383 & -2.120 & \nodata          & $(    6.2, 7.8)$ & B     &  Y?  &  Y \\
W-9   & 190.481 & -2.054 & $(   -5.4,-3.0)$ & $(    5.6, 7.6)$ & C$|$C &  Y?  &  Y \\
W-10  & 190.529 & -2.358 & $(   -4.8,-2.2)$ & \nodata          & C     &  Y?  &  Y \\
W-11  & 191.948 & -2.520 & \nodata          & $(    5.0, 6.3)$ & B     &  Y?   &  Y \\
W-12  & 192.477 & -2.725 & $(   -1.0, 1.5)$ & $(    4.0, 5.9)$ & B$|$B &  Y?  &  Y \\
W-13  & 192.721 & -2.538 & \nodata          & $(    5.3, 7.6)$ & C     &  Y?  &  Y \\
W-14  & 193.008 & -2.461 & \nodata          & $(    5.0, 7.8)$ & B     &  Y?   &  Y \\
W-15  & 193.021 & -2.575 & \nodata          & $(    5.1, 6.8)$ & C     &  Y?  &  Y \\
W-16  & 193.592 & -3.075 & $(    0.0, 1.0)$ & \nodata          & B     &  Y?   &  Y \\
W-17  & 193.985 & -3.380 & $(   -0.2, 1.8)$ & \nodata          & B     &  Y?   &  Y \\
W-18  & 194.083 & -2.880 & $(   -1.1, 0.2)$ & \nodata          & B     &  Y?  &  Y \\
W-19  & 194.183 & -2.836 & $(   -0.8, 0.3)$ & $(    4.0, 6.0)$ & B$|$C &  Y   &  Y \\
\hline
\multicolumn{8}{l}{Swallow} \\
\hline
S-1   & 193.858 & -0.600 & \nodata          & $(   13.9,16.2)$ & B     &  Y   &  Y \\
S-2   & 194.275 & -0.965 & $(   10.6,11.6)$ & \nodata          & B     &  Y?  &  Y \\
S-3   & 194.710 & -1.163 & \nodata          & $(   17.0,19.0)$ & B     &  Y   &  Y \\
S-4   & 194.840 & -1.155 & \nodata          & $(   18.0,19.8)$ & C     &  Y?  &  Y \\
S-5   & 194.920 & -1.175 & \nodata          & $(   18.4,20.4)$ & B     &  Y?  &  Y \\
S-6   & 194.935 & -1.223 & $(    8.7,13.0)$ & $(   18.8,23.8)$ & A$|$A &  Y   &  Y \\
S-7   & 194.975 & -1.340 & $(   12.4,14.0)$ & \nodata          & B     &  Y   &  Y \\
S-8   & 195.003 & -1.569 & $(    9.5,12.0)$ & $(   14.6,16.0)$ & A$|$A &  Y   &  Y \\
S-9   & 195.010 & -1.915 & \nodata          & $(   16.0,17.1)$ & B     &  Y?   &  Y \\
S-10  & 195.021 & -1.370 & $(   11.5,13.0)$ & \nodata          & B     &  Y?   &  Y \\
\hline
\multicolumn{8}{l}{Horn} \\
\hline
H-1   & 188.587 & -1.496 & $(   11.6,13.0)$ & $(   16.5,17.5)$ & B$|$B &  Y?  &  Y \\
H-2   & 188.647 & -1.786 & $(    8.5,11.5)$ & $(   14.3,18.3)$ & A$|$C &  Y   &  Y \\
H-3   & 188.807 & -1.443 & $(   11.7,13.6)$ & $(   16.6,18.9)$ & B$|$B &  Y   &  Y \\
H-4   & 188.931 & -1.427 & $(   13.0,14.3)$ & $(   17.0,19.4)$ & A$|$B &  Y   &  Y \\
H-5   & 189.083 & -1.523 & $(   11.7,13.0)$ & $(   18.0,19.4)$ & B$|$C &  Y?  &  Y \\
H-6   & 189.115 & -1.495 & \nodata          & $(   18.5,19.7)$ & A     &  Y?   &  Y \\
H-7   & 189.219 & -1.065 & $(    9.2,12.0)$ & $(   16.3,17.9)$ & A$|$B &  Y   &  Y \\
H-8   & 189.467 & -1.242 & $(   10.5,13.8)$ & $(   18.3,22.1)$ & A$|$A &  Y   &  Y \\
H-9   & 189.875 & -1.363 & $(   14.0,15.2)$ & $(   17.4,19.8)$ & B$|$A &  Y   &  Y \\
\hline
\multicolumn{8}{l}{Remote} \\
\hline
R-1   & 192.544 & -0.154 & $(   21.3,23.1)$ & $(   25.5,27.1)$ & B$|$A &  Y   &  Y \\
R-2   & 192.837 &  0.617 & \nodata          & $(   21.1,22.2)$ & B     &  Y?  &  Y \\
R-3   & 192.915 & -0.629 & $(   19.2,21.1)$ & $(   23.9,25.7)$ & A$|$A &  Y   &  Y \\
R-4   & 192.962 & -0.513 & $(   20.0,21.1)$ & $(   22.9,24.3)$ & A$|$B &  Y?  &  Y \\
R-5   & 193.433 & -1.198 & \nodata          & $(   27.5,30.0)$ & A     &  Y   &  Y \\
R-6   & 193.534 & -1.107 & $(   20.3,21.8)$ & $(   24.9,26.0)$ & C$|$B &  Y?  &  Y \\
R-7   & 193.695 & -1.060 & $(   19.5,20.8)$ & \nodata          & B     &  Y   &  Y \\
R-8   & 193.721 & -1.113 & $(   19.8,21.3)$ & \nodata          & B     &  Y?  &  Y \\
R-9   & 193.792 &  1.138 & $(   18.6,19.7)$ & \nodata          & B     &  Y?  &  Y \\
R-10  & 193.795 &  0.983 & $(   17.5,18.6)$ & \nodata          & A     &  Y?  &  Y \\
R-11  & 193.825 & -0.255 & $(   17.0,19.0)$ & $(   22.2,23.5)$ & B$|$C &  Y   &  Y \\
R-12  & 193.840 &  0.945 & $(   18.0,19.0)$ & \nodata          & B     &  Y?  &  Y \\
R-13  & 193.850 & -0.442 & \nodata          & $(   21.8,23.2)$ & C     &  Y?   &  Y \\
R-14  & 193.858 &  1.129 & \nodata          & $(   22.5,23.5)$ & B     &  Y?  &  Y \\
R-15  & 193.875 &  0.818 & $(   16.0,18.0)$ & $(   21.0,22.0)$ & C$|$C &  Y?  &  Y \\
R-16  & 193.900 &  1.054 & \nodata          & $(   24.2,26.0)$ & A     &  Y   &  Y \\
R-17  & 193.985 & -1.029 & $(   21.0,22.7)$ & \nodata          & C     &  Y?   &  Y \\
R-18  & 194.004 & -0.310 & $(   17.0,18.5)$ & $(   21.6,23.3)$ & C$|$C &  Y?  &  Y \\
R-19  & 194.008 &  0.836 & $(   17.0,19.0)$ & $(   21.3,22.5)$ & B$|$B &  Y?  &  Y \\
R-20  & 194.062 & -0.192 & \nodata          & $(   21.0,22.5)$ & B     &  Y   &  Y \\
R-21  & 194.162 & -0.117 & $(   16.3,18.3)$ & \nodata          & C     &  Y   &  Y \\
R-22  & 194.440 &  0.045 & $(   15.2,17.8)$ & $(   20.8,23.2)$ & B$|$B &  Y   &  Y \\
R-23  & 194.450 & -0.867 & \nodata          & $(   25.7,27.3)$ & A     &  Y   &  Y \\
R-24  & 194.479 & -0.010 & $(   14.3,17.1)$ & \nodata          & A     &  Y?  &  Y \\
\enddata
\tablenotetext{a}{\citet{WWZ2004}, and references therein, $^\mathrm{b}$\citet{BL1983}, $^\mathrm{c}$\citet{SHD1988}}
\tablecomments{Y = Yes, Y? = possible, N = Not a new detection. Because the outflow candidates with score D were removed from the final sample (see Appendix A. 6), we did not include them in this table.}
\end{deluxetable}

\clearpage

\section{The Physical Properties of Outflow Samples}

This appendix presents the physical properties of outflow samples, where the calculation of those physical parameters are presented in Section 4.3 and Appendix B. The definition of the location of outflow candidates refers to Appendix A. 5. All physical parameters presented here are not corrected. The distances of Swallow, Horn and remote are refer to Paper I.

\begin{deluxetable}{llrrrrrrrrr}
\centering
\setlength\tabcolsep{3pt}
\tablecolumns{11}
\tabletypesize{\small}
\tablewidth{0pt}
\tablenum{D}
\tablecaption{Physical Properties of the Outflow Samples}
\tablehead{
\colhead{Index} & \colhead{Lobe} & \colhead{$l$} & \colhead{$b$} & \colhead{$\langle\Delta v_{\mathrm{lobe}}\rangle$} & \colhead{$l_{\mathrm{lobe}}$} & \colhead{$M_{\mathrm{lobe}}$} & \colhead{$P_{\mathrm{lobe}}$} & \colhead{$E_{\mathrm{lobe}}$} & \colhead{$t_d$} & \colhead{$L_{\mathrm{m}}$} \\
 \colhead{} & \colhead{} & \colhead{($\degr$)} &  \colhead{($\degr$)} & \colhead{(km s$^{-1}$)} & \colhead{(pc)} & \colhead{(M$_{\odot}$)} & \colhead{(M$_{\odot}$ km s$^{-1}$)} & \colhead{(10$^{43}$ erg)} & \colhead{(10$^5$ yr)} & \colhead{(10$^{30}$ erg s$^{-1}$)}
}
\startdata
\multicolumn{11}{l}{GGMC 1} \\
\hline
G1-1     &   Red &  192.720 &  -1.245  &   2.2 &  2.66  &     0.60   &    1.29   &     2.77      &   8.9       &     1.0    \\
G1-2     &   Red &  192.842 &  -1.133  &   2.4 &  1.71  &     0.43   &    1.06   &     2.57      &   5.2       &     1.6    \\
G1-3     &   Red &  193.093 &  -1.388  &   2.7 &  1.80  &     0.38   &    1.05   &     2.83      &   4.2       &     2.1    \\
G1-4     &   Blue &  193.235 &  -1.620  &  1.7 &  1.23   &     0.24   &    0.41   &     0.70      &  5.2        &    0.4     \\
         &   Red &  193.242 &  -1.618  &   1.9 &  1.33  &     0.26   &    0.50   &     0.93      &   4.8       &     0.6    \\
G1-5     &   Red &  193.250 &  -1.038  &   2.7 &  3.93  &     1.03   &    2.81   &     7.58      &   11.4      &     2.1    \\
G1-6     &   Blue &  193.310 &  -1.368  &  2.8 &  1.42   &     0.52   &    1.46   &     4.03      &  3.8        &    3.3     \\
         &   Red &  193.310 &  -1.380  &   3.0 &  1.24  &     0.62   &    1.88   &     5.60      &   2.7       &     6.6    \\
G1-7     &   Blue &  193.505 &  -1.150  &  2.5 &  2.74   &     1.79   &    4.44   &     10.90     &  8.3        &    4.1     \\
         &   Red &  193.490 &  -1.160  &   2.4 &  2.02  &     1.22   &    2.91   &     6.88      &   5.4       &     4.1    \\
G1-8     &   Red &  193.506 &  -0.029  &   1.7 &  2.02  &     0.49   &    0.85   &     1.47      &   8.0       &     0.6    \\
G1-9     &   Blue &  193.617 &  -0.138  &  2.8 &  1.33   &     0.25   &    0.71   &     1.99      &  3.9        &    1.6     \\
G1-10    &   Blue &  193.650 &  -0.610  &  2.3 &  1.85   &     0.79   &    1.79   &     4.02      &  6.2        &    2.1     \\
G1-11    &   Red &  193.680 &  0.185   &   1.3 &  2.54  &     0.33   &    0.42   &     0.54      &   13.3      &     0.1    \\
G1-12    &   Blue &  193.717 &  -0.275  &  3.0 &  1.85   &     0.88   &    2.64   &     7.87      &  4.5        &    5.5     \\
G1-13    &   Blue &  193.767 &  -0.800  &  2.3 &  1.42   &     0.24   &    0.54   &     1.21      &  4.9        &    0.8     \\
G1-14    &   Blue &  193.845 &  -0.395  &  2.3 &  2.37   &     1.74   &    4.04   &     9.32      &  7.5        &    4.0     \\
G1-15    &   Blue &  193.885 &  -1.086  &  4.2 &  1.25   &     0.46   &    1.96   &     8.24      &  2.3        &    11.9    \\
G1-16    &   Blue &  193.908 &  0.205   &  2.1 &  1.34   &     0.15   &    0.32   &     0.66      &  4.8        &    0.4     \\
G1-17    &   Blue &  193.915 &  -1.000  &  2.1 &  1.88   &     0.87   &    1.85   &     3.90      &  6.5        &    1.9     \\
G1-18    &   Blue &  193.998 &  0.238   &  1.6 &  1.95   &     0.58   &    0.92   &     1.46      &  9.0        &    0.5     \\
G1-19    &   Blue &  194.017 &  -0.729  &  2.8 &  1.48   &     0.31   &    0.87   &     2.40      &  4.2        &    1.8     \\
G1-20    &   Red &  194.020 &  -1.095  &   3.3 &  2.21  &     0.91   &    2.99   &     9.76      &   4.8       &     6.5    \\
G1-21    &   Red &  194.025 &  -0.875  &   2.8 &  2.33  &     1.49   &    4.19   &     11.70     &   5.4       &     7.0    \\
G1-22    &   Blue &  194.054 &  -0.671  &  3.2 &  2.41   &     1.20   &    3.89   &     12.50     &  5.6        &    7.0     \\
         &   Red &  194.058 &  -0.655  &   2.4 &  2.23  &     0.51   &    1.23   &     2.94      &   6.6       &     1.4    \\
G1-23    &   Blue &  194.062 &  -0.238  &  2.3 &  1.64   &     0.66   &    1.52   &     3.48      &  5.2        &    2.1     \\
G1-24    &   Red &  194.110 &  -0.485  &   1.6 &  3.34  &     2.26   &    3.72   &     6.07      &   12.5      &     1.6    \\
G1-25    &   Red &  194.150 &  -0.792  &   2.8 &  2.26  &     0.82   &    2.33   &     6.53      &   5.2       &     4.0    \\
G1-26    &   Blue &  194.175 &  -0.165  &  1.8 &  2.20   &     0.73   &    1.28   &     2.23      &  8.7        &    0.9     \\
G1-27    &   Red &  194.180 &  -1.100  &   2.8 &  1.34  &     0.41   &    1.16   &     3.24      &   3.2       &     3.1    \\
G1-28    &   Red &  194.185 &  -0.935  &   2.2 &  3.54  &     3.21   &    7.15   &     15.80     &   9.9       &     5.0    \\
G1-29    &   Blue &  194.235 &  0.113   &  3.0 &  2.72   &     1.38   &    4.17   &     12.50     &  6.9        &    5.8     \\
G1-30    &   Blue &  194.246 &  -0.196  &  1.7 &  1.40   &     0.18   &    0.30   &     0.49      &  6.2        &    0.3     \\
G1-31    &   Blue &  194.310 &  -0.263  &  2.3 &  4.00   &     2.01   &    4.53   &     10.10     &  12.4       &    2.6     \\
G1-32    &   Blue &  194.335 &  -0.990  &  2.2 &  2.43   &     0.91   &    1.98   &     4.29      &  8.5        &    1.6     \\
G1-33    &   Blue &  194.367 &  -0.790  &  2.2 &  1.58   &     0.48   &    1.06   &     2.33      &  5.5        &    1.3     \\
         &   Red &  194.373 &  -0.790  &   1.5 &  2.02  &     0.47   &    0.73   &     1.12      &   9.3       &     0.4    \\
G1-34    &   Blue &  194.672 &  0.318   &  3.0 &  2.48   &     1.08   &    3.22   &     9.50      &  6.2        &    4.8     \\
         &   Red &  194.650 &  0.320   &   2.4 &  1.30  &     0.45   &    1.08   &     2.54      &   3.5       &     2.3    \\
\hline
\multicolumn{11}{l}{GGMC 2} \\
\hline
G2-1     &   Blue &  192.455 &  0.005   & 3.0  & 2.75    &     0.99   &    2.99   &     8.97      &  6.8        &    4.2     \\
G2-2     &   Red &  192.580 &  0.210   &  1.9  & 1.34   &     0.65   &    1.24   &     2.32      &   4.1       &     1.8    \\
G2-3     &   Blue &  192.604 &  -0.054  & 5.4  & 1.49    &     1.19   &    6.44   &     34.50     &  2.0        &    53.9    \\
         &   Red &  192.589 &  -0.042  &  5.6  & 2.10   &     3.97   &    22.30  &     124.00    &   2.4       &     164.0  \\
G2-4     &   Blue &  192.608 &  -0.129  & 4.8  & 1.49    &     0.92   &    4.39   &     20.80     &  2.4        &    28.4    \\
G2-5     &   Red &  192.655 &  -0.210  &  4.2  & 1.83   &     0.84   &    3.53   &     14.70     &   3.4       &     14.1   \\
G2-6     &   Blue &  192.695 &  0.193   & 2.2  & 2.06    &     0.85   &    1.88   &     4.12      &  7.5        &    1.8     \\
         &   Red &  192.695 &  0.210   &  1.7  & 1.99   &     1.02   &    1.69   &     2.76      &   7.8       &     1.1    \\
G2-7     &   Blue &  192.720 &  -0.233  & 4.0  & 2.12    &     1.29   &    5.16   &     20.40     &  4.1        &    16.1    \\
         &   Red &  192.710 &  -0.275  &  3.5  & 3.48   &     6.85   &    24.10  &     83.80     &   7.2       &     37.3   \\
G2-8     &   Red &  192.718 &  0.025   &  3.0  & 1.73   &     1.74   &    5.18   &     15.30     &   3.7       &     13.1   \\
G2-9     &   Blue &  192.730 &  -0.070  & 3.3  & 2.31    &     1.73   &    5.69   &     18.50     &  5.4        &    10.9    \\
G2-10    &   Blue &  192.783 &  -0.195  & 3.7  & 1.52    &     0.90   &    3.35   &     12.40     &  3.1        &    12.6    \\
         &   Red &  192.758 &  -0.200  &  4.1  & 1.28   &     0.28   &    1.18   &     4.82      &   2.1       &     7.1    \\
G2-11    &   Red &  192.820 &  0.125   &  3.5  & 1.61   &     0.65   &    2.30   &     8.05      &   3.1       &     8.2    \\
G2-12    &   Red &  192.846 &  0.288   &  2.2  & 2.10   &     1.06   &    2.33   &     5.08      &   6.5       &     2.5    \\
G2-13    &   Red &  192.875 &  -0.060  &  2.8  & 1.83   &     1.51   &    4.15   &     11.30     &   4.4       &     8.1    \\
G2-14    &   Blue &  192.950 &  0.195   & 4.3  & 2.26    &     1.81   &    7.73   &     32.70     &  4.1        &    25.8    \\
         &   Red &  192.914 &  0.190   &  3.0  & 4.53   &     8.37   &    24.90  &     73.20     &   9.3       &     25.1   \\
G2-15    &   Red &  192.979 &  0.092   &  4.5  & 1.51   &     1.56   &    7.02   &     31.30     &   1.8       &     53.2   \\
G2-16    &   Blue &  192.992 &  0.158   & 6.5  & 1.48    &     1.27   &    8.32   &     54.00     &  1.6        &    107.2   \\
         &   Red &  192.992 &  0.167   &  4.4  & 2.61   &     3.03   &    13.20  &     56.70     &   3.8       &     46.7   \\
G2-17    &   Blue &  193.090 &  0.208   & 2.8  & 1.48    &     0.51   &    1.41   &     3.87      &  4.1        &    3.0     \\
G2-18    &   Red &  193.197 &  0.254   &  1.6  & 1.80   &     0.63   &    0.99   &     1.53      &   7.1       &     0.7    \\
G2-19    &   Blue &  193.290 &  0.200   & 2.6  & 1.35    &     0.62   &    1.60   &     4.08      &  3.9        &    3.3     \\
         &   Red &  193.287 &  0.171   &  1.5  & 5.65   &     9.23   &    14.10  &     21.30     &   22.3      &     3.1    \\
G2-20    &   Red &  193.320 &  0.030   &  1.8  & 2.58   &     0.81   &    1.45   &     2.59      &   10.0      &     0.9    \\
\hline
\multicolumn{11}{l}{BFS 52} \\
\hline
B-1      &   Blue &  191.067 &  0.810   &  3.8 & 1.21    &     0.49   &    1.88   &     7.10      & 2.4        &    9.7     \\
         &   Red &  191.065 &  0.827   &   2.5 & 1.76   &     0.56   &    1.41   &     3.47      &  4.8       &     2.3    \\
B-2      &   Red &  191.083 &  0.740   &   1.5 & 1.34   &     0.19   &    0.28   &     0.42      &  6.2       &     0.2    \\
B-3      &   Red &  191.137 &  0.662   &   2.3 & 3.07   &     1.20   &    2.76   &     6.30      &  8.0       &     2.5    \\
B-4      &   Red &  191.220 &  0.752   &   1.9 & 1.66   &     0.61   &    1.14   &     2.13      &  5.2       &     1.3    \\
B-5      &   Blue &  191.300 &  0.979   &  1.9 & 2.38    &     0.71   &    1.36   &     2.56      & 10.0       &    0.9     \\
         &   Red &  191.308 &  0.933   &   1.6 & 1.41   &     0.37   &    0.57   &     0.88      &  5.9       &     0.5    \\
B-6      &   Red &  191.500 &  0.615   &   1.5 & 2.90   &     1.94   &    2.98   &     4.53      &  12.3      &     1.2    \\
B-7      &   Blue &  191.742 &  0.833   &  2.3 & 1.85    &     0.92   &    2.08   &     4.65      & 5.9        &    2.5     \\
B-8      &   Blue &  191.790 &  0.783   &  1.6 & 1.76    &     0.66   &    1.04   &     1.60      & 8.7        &    0.6     \\
         &   Red &  191.803 &  0.786   &   2.0 & 1.73   &     0.90   &    1.84   &     3.71      &  5.5       &     2.1    \\
B-9      &   Red &  191.827 &  1.000   &   1.7 & 1.51   &     0.33   &    0.58   &     1.00      &  5.6       &     0.6    \\
B-10     &   Blue &  191.895 &  0.778   &  1.9 & 2.09    &     0.46   &    0.85   &     1.56      & 8.9        &    0.6     \\
B-11     &   Blue &  191.892 &  1.115   &  2.2 & 1.80    &     0.40   &    0.88   &     1.93      & 6.2        &    1.0     \\
         &   Red &  191.900 &  1.075   &   1.9 & 1.52   &     0.31   &    0.58   &     1.08      &  6.1       &     0.6    \\
B-12     &   Blue &  191.893 &  0.880   &  2.2 & 2.88    &     3.19   &    6.98   &     15.10     & 9.6        &    5.0     \\
         &   Red &  191.917 &  0.908   &   3.6 & 1.86   &     1.62   &    5.82   &     20.80     &  3.1       &     20.9   \\
B-13     &   Blue &  192.030 &  1.050   &  1.8 & 1.41    &     0.24   &    0.43   &     0.74      & 6.1        &    0.4     \\
         &   Red &  192.030 &  1.042   &   1.8 & 1.42   &     0.30   &    0.54   &     0.94      &  5.1       &     0.6    \\
B-14     &   Blue &  192.742 &  1.282   &  1.2 & 1.38    &     0.29   &    0.35   &     0.42      & 8.2        &    0.1     \\
B-15     &   Blue &  193.009 &  1.067   &  1.5 & 1.24    &     0.16   &    0.24   &     0.35      & 6.3        &    0.1     \\
B-16     &   Blue &  193.167 &  0.695   &  3.3 & 2.03    &     0.33   &    1.09   &     3.57      & 4.9        &    2.3     \\
         &   Red &  193.110 &  0.705   &   1.4 & 1.58   &     0.30   &    0.41   &     0.55      &  7.5       &     0.2    \\
B-17     &   Blue &  193.145 &  0.948   &  2.2 & 2.03    &     0.53   &    1.19   &     2.63      & 7.3        &    1.1     \\
         &   Red &  193.177 &  0.940   &   2.0 & 1.68   &     0.38   &    0.77   &     1.54      &  5.6       &     0.9    \\
\hline
\multicolumn{11}{l}{GGMC 3} \\
\hline
G3-1     &   Red &  189.221 &  0.846   &  4.6 &  1.72   &     0.78   &    3.57   &     16.20     &  2.7      &    18.7   \\
G3-2     &   Red &  189.225 &  0.767   &  1.8 &  1.89   &     0.34   &    0.59   &     1.03      &  7.6      &    0.4    \\
G3-3     &   Blue &  189.237 &  -0.529  & 1.9 &  1.80    &     0.39   &    0.73   &     1.35      & 7.3       &   0.6     \\
G3-4     &   Blue &  189.354 &  -0.413  & 2.8 &  2.37    &     0.99   &    2.76   &     7.63      & 6.2       &   4.0     \\
G3-5     &   Red &  189.383 &  0.463   &  3.8 &  1.23   &     0.35   &    1.32   &     4.91      &  2.1      &    7.3    \\
G3-6     &   Red &  189.580 &  -0.145  &  2.5 &  1.69   &     0.59   &    1.46   &     3.61      &  4.8      &    2.4    \\
G3-7     &   Red &  189.583 &  0.000   &  4.6 &  1.54   &     0.71   &    3.24   &     14.70     &  2.1      &    21.4   \\
G3-8     &   Blue &  189.630 &  0.190   & 5.1 &  1.86    &     1.35   &    6.93   &     35.30     & 3.1       &   36.4    \\
         &   Red &  189.587 &  0.150   &  3.5 &  2.27   &     1.00   &    3.50   &     12.10     &  4.9      &    7.8    \\
G3-9     &   Red &  189.633 &  0.085   &  3.6 &  2.48   &     1.57   &    5.63   &     20.00     &  5.1      &    12.6   \\
G3-10    &   Blue &  189.704 &  0.067   & 5.2 &  1.30    &     0.53   &    2.75   &     14.10     & 2.0       &   22.2    \\
G3-11    &   Blue &  189.742 &  0.025   & 4.5 &  1.86    &     0.28   &    1.25   &     5.51      & 3.5       &   5.0     \\
         &   Red &  189.717 &  0.021   &  4.5 &  1.85   &     0.69   &    3.11   &     14.00     &  3.2      &    13.6   \\
G3-12    &   Red &  189.742 &  -0.040  &  3.1 &  2.04   &     0.84   &    2.58   &     7.92      &  4.7      &    5.5    \\
G3-13    &   Blue &  189.747 &  0.165   & 4.8 &  2.23    &     1.65   &    7.82   &     36.90     & 3.7       &   32.0    \\
G3-14    &   Blue &  189.804 &  0.346   & 6.9 &  2.86    &     7.67   &    53.20  &     365.00    & 3.2       &   357.8   \\
         &   Red &  189.777 &  0.342   &  5.0 &  1.96   &     4.18   &    20.80  &     103.00    &  2.5      &    127.1  \\
G3-15    &   Red &  189.792 &  0.467   &  4.0 &  2.79   &     2.75   &    11.00  &     43.50     &  4.9      &    27.8   \\
G3-16    &   Red &  189.804 &  -0.625  &  2.1 &  1.93   &     0.44   &    0.91   &     1.87      &  6.5      &    0.9    \\
G3-17    &   Blue &  189.821 &  0.088   & 5.8 &  3.71    &     5.47   &    31.90  &     184.00    & 5.1       &   115.7   \\
G3-18    &   Red &  189.833 &  0.313   &  5.3 &  1.44   &     2.04   &    10.90  &     57.20     &  1.8      &    98.7   \\
G3-19    &   Blue &  189.852 &  0.167   & 5.4 &  1.85    &     0.66   &    3.61   &     19.40     & 2.8       &   21.4    \\
         &   Red &  189.833 &  0.221   &  5.7 &  1.65   &     0.38   &    2.14   &     12.00     &  2.4      &    16.3   \\
G3-20    &   Blue &  189.942 &  0.225   & 3.7 &  2.40    &     0.88   &    3.20   &     11.60     & 5.4       &   6.9     \\
G3-21    &   Blue &  189.992 &  0.340   & 5.8 &  1.75    &     3.32   &    19.10  &     109.00    & 2.3       &   154.8   \\
         &   Red &  189.954 &  0.321   &  4.5 &  1.47   &     1.81   &    8.09   &     35.90     &  2.0      &    56.2   \\
G3-22    &   Red &  190.008 &  -0.138  &  4.2 &  3.13   &     1.08   &    4.52   &     18.80     &  5.2      &    11.4   \\
G3-23    &   Blue &  190.017 &  0.125   & 4.2 &  3.71    &     4.55   &    19.00  &     78.80     & 7.2       &   35.1    \\
G3-24    &   Blue &  190.030 &  -0.665  & 3.0 &  1.89    &     0.65   &    1.94   &     5.72      & 4.8       &   3.8     \\
G3-25    &   Blue &  190.040 &  0.520   & 3.4 &  2.16    &     4.25   &    14.60  &     49.80     & 4.5       &   34.6    \\
         &   Red &  190.050 &  0.525   &  2.6 &  2.12   &     0.74   &    1.93   &     5.00      &  6.1      &    2.6    \\
G3-26    &   Blue &  190.075 &  0.225   & 5.1 &  2.52    &     1.83   &    9.27   &     46.50     & 3.9       &   37.3    \\
         &   Red &  190.087 &  0.267   &  3.4 &  2.26   &     0.73   &    2.44   &     8.10      &  5.2      &    5.0    \\
G3-27    &   Blue &  190.067 &  0.650   & 4.6 &  1.25    &     0.62   &    2.83   &     12.80     & 2.1       &   19.9    \\
         &   Red &  190.121 &  0.675   &  4.3 &  1.41   &     0.26   &    1.12   &     4.80      &  2.5      &    6.2    \\
G3-28    &   Blue &  190.090 &  0.615   & 3.0 &  1.99    &     3.03   &    8.95   &     26.20     & 5.2       &   16.0    \\
         &   Red &  190.112 &  0.592   &  2.5 &  2.17   &     1.01   &    2.51   &     6.19      &  6.1      &    3.3    \\
G3-29    &   Blue &  190.175 &  0.725   & 4.7 &  1.33    &     0.38   &    1.78   &     8.26      & 2.4       &   11.2    \\
         &   Red &  190.158 &  0.746   &  2.6 &  2.76   &     2.00   &    5.18   &     13.30     &  7.2      &    5.8    \\
G3-30    &   Blue &  190.179 &  0.175   & 3.8 &  1.72    &     0.92   &    3.50   &     13.30     & 3.4       &   12.4    \\
         &   Red &  190.182 &  0.170   &  3.2 &  1.38   &     0.45   &    1.45   &     4.64      &  3.2      &    4.5    \\
G3-31    &   Red &  190.185 &  0.365   &  1.8 &  1.52   &     0.45   &    0.81   &     1.44      &  5.5      &    0.9    \\
G3-32    &   Blue &  190.254 &  0.900   & 3.3 &  1.79    &     1.47   &    4.88   &     16.00     & 3.9       &   13.1    \\
         &   Red &  190.257 &  0.900   &  2.5 &  1.75   &     1.58   &    3.87   &     9.39      &  4.5      &    6.6    \\
G3-33    &   Red &  190.275 &  0.365   &  1.7 &  2.37   &     0.92   &    1.52   &     2.49      &  8.2      &    1.0    \\
G3-34    &   Blue &  190.337 &  0.046   & 2.9 &  1.82    &     0.71   &    2.09   &     6.10      & 4.5       &   4.3     \\
G3-35    &   Blue &  190.410 &  0.960   & 2.7 &  2.82    &     1.41   &    3.78   &     10.00     & 7.8       &   4.1     \\
G3-36    &   Blue &  190.415 &  0.620   & 1.8 &  2.79    &     2.58   &    4.50   &     7.79      & 10.7      &   2.3     \\
G3-37    &   Blue &  190.546 &  1.100   & 2.1 &  2.04    &     0.47   &    0.97   &     2.01      & 7.5       &   0.9     \\
G3-38    &   Blue &  190.558 &  0.813   & 4.2 &  1.64    &     1.10   &    4.59   &     19.00     & 2.8       &   21.9    \\
G3-39    &   Red &  190.562 &  0.600   &  1.4 &  2.81   &     1.76   &    2.52   &     3.56      &  11.4     &    1.0    \\
G3-40    &   Red &  190.567 &  0.743   &  2.9 &  4.58   &     11.40  &    33.40  &     96.60     &  8.9      &    34.6   \\
G3-41    &   Red &  190.692 &  0.425   &  2.0 &  2.82   &     1.44   &    2.93   &     5.91      &  8.6      &    2.2    \\
G3-42    &   Blue &  190.738 &  0.805   & 2.4 &  1.64    &     0.49   &    1.18   &     2.80      & 5.4       &   1.7     \\
G3-43    &   Red &  190.775 &  0.696   &  6.7 &  1.21   &     0.26   &    1.69   &     11.10     &  1.6      &    22.4   \\
G3-44    &   Blue &  190.937 &  0.308   & 3.1 &  1.58    &     0.58   &    1.77   &     5.35      & 3.9       &   4.3     \\
         &   Red &  190.900 &  0.333   &  1.8 &  1.21   &     0.16   &    0.29   &     0.52      &  4.8      &    0.4    \\
G3-45    &   Blue &  191.045 &  0.351   & 1.7 &  1.33    &     0.74   &    1.25   &     2.11      & 5.8       &   1.2     \\
         &   Red &  191.060 &  0.370   &  2.4 &  1.82   &     0.77   &    1.80   &     4.19      &  5.5      &    2.4    \\
\hline
\multicolumn{11}{l}{GGMC 4} \\
\hline
G4-1     &   Red &  188.673 &  0.935   &   1.8 & 2.45   &     0.90   &    1.65   &     2.99      &   9.4      &    1.0    \\
G4-2     &   Blue &  188.790 &  1.290   &  1.9 & 2.13    &     0.63   &    1.22   &     2.33      &  8.2       &   0.9     \\
G4-3     &   Blue &  188.796 &  1.026   &  3.9 & 1.49    &     0.70   &    2.74   &     10.70     &  3.1       &   10.9    \\
G4-4     &   Blue &  188.812 &  1.063   &  4.7 & 1.71    &     0.71   &    3.28   &     15.10     &  3.0       &   16.5    \\
G4-5     &   Red &  188.817 &  0.933   &   2.2 & 1.48   &     0.34   &    0.77   &     1.70      &   4.8      &    1.1    \\
G4-6     &   Blue &  188.940 &  1.060   &  4.2 & 2.16    &     1.49   &    6.32   &     26.50     &  4.2       &   20.0    \\
G4-7    &   Blue &  188.950 &  0.883   &  7.1 &  1.99   &     3.52   &    24.90  &     175.00    &   2.1      &    265.5  \\
         &   Red &  188.937 &  0.883   &   6.4 & 1.64   &     3.76   &    24.00  &     152.00    &   1.4      &    329.4  \\
G4-8     &   Blue &  188.979 &  1.125   &  5.2 & 1.86    &     1.67   &    8.66   &     44.60     &  2.7       &   52.3    \\
G4-9     &   Blue &  189.000 &  1.213   &  6.0 & 1.73    &     0.65   &    3.87   &     23.00     &  2.4       &   31.3    \\
G4-10    &   Blue &  189.037 &  0.808   &  6.8 & 1.51    &     2.03   &    13.80  &     92.30     &  1.6       &   182.5   \\
         &   Red &  189.042 &  0.792   &   5.3 & 1.96   &     3.32   &    17.50  &     91.10     &   2.4      &    122.8  \\
G4-11    &   Blue &  189.058 &  1.079   &  4.9 & 1.42    &     1.08   &    5.30   &     25.80     &  2.3       &   37.1    \\
G4-12    &   Blue &  189.160 &  0.455   &  1.1 & 1.88    &     0.32   &    0.36   &     0.40      &  12.1      &   0.1     \\
G4-13    &   Blue &  189.222 &  0.355   &  1.3 & 1.25    &     0.28   &    0.37   &     0.48      &  7.2       &   0.2     \\
         &   Red &  189.180 &  0.391   &   1.4 & 1.79   &     0.37   &    0.53   &     0.75      &   7.8      &    0.3    \\
G4-14    &   Red &  189.268 &  1.285   &   5.9 & 2.33   &     0.63   &    3.74   &     21.90     &   3.2      &    21.2   \\
G4-15    &   Blue &  189.400 &  1.783   &  2.0 & 1.76    &     0.50   &    1.02   &     2.06      &  6.1       &   1.1     \\
G4-16    &   Red &  189.454 &  0.738   &   2.6 & 2.14   &     0.65   &    1.72   &     4.49      &   5.8      &    2.5    \\
G4-17    &   Red &  189.575 &  0.968   &   2.4 & 1.55   &     0.51   &    1.23   &     2.92      &   4.4      &    2.1    \\
G4-18    &   Blue &  189.687 &  1.779   &  2.5 & 1.76    &     0.27   &    0.68   &     1.70      &  5.2       &   1.1     \\
\hline
\multicolumn{11}{l}{Local Lynds} \\
\hline
L-1      &   Blue &  190.660 &  -0.400  &  2.2 & 0.44    &     0.03   &    0.06   &     0.14      &  1.4     &    0.3     \\
L-2      &   Blue &  192.028 &  -0.956  &  4.8 & 0.28    &     0.02   &    0.09   &     0.42      &  0.4     &    2.8     \\
         &   Red &  192.020 &  -0.973  &   2.4 & 0.58   &     0.06   &    0.15   &     0.37      &   1.7    &     0.7    \\
\hline
\multicolumn{11}{l}{West Front} \\
\hline
W-1      &   Blue &  187.092 &  -0.352  &  2.3  & 0.97   &     0.08   &    0.18   &     0.39      &  3.4     &    0.4     \\
W-2      &   Blue &  187.535 &  -0.728  &  3.2  & 0.76   &     0.16   &    0.51   &     1.61      &  1.7     &    3.1     \\
W-3      &   Blue &  187.562 &  -3.215  &  2.8  & 0.55   &     0.04   &    0.11   &     0.30      &  1.4     &    0.6     \\
         &   Red &  187.562 &  -3.207  &   2.3  & 0.59  &     0.03   &    0.08   &     0.18      &   2.0    &     0.3    \\
W-4      &   Blue &  188.275 &  -2.875  &  1.2  & 0.52   &     0.05   &    0.05   &     0.06      &  3.2     &    0.1     \\
         &   Red &  188.267 &  -2.862  &   1.8  & 0.89  &     0.25   &    0.46   &     0.82      &   2.8    &     0.9    \\
W-5      &   Blue &  188.970 &  -1.940  &  4.1  & 0.37   &     0.05   &    0.19   &     0.76      &  0.7     &    3.6     \\
W-6      &   Blue &  189.244 &  -1.583  &  2.7  & 0.44   &     0.05   &    0.13   &     0.36      &  1.3     &    0.9     \\
W-7      &   Red &  190.133 &  -2.420  &   5.6  & 0.97  &     0.24   &    1.36   &     7.58      &   1.4    &     17.7   \\
W-8      &   Red &  190.383 &  -2.120  &   5.2  & 0.62  &     0.05   &    0.24   &     1.20      &   1.0    &     3.8    \\
W-9      &   Blue &  190.450 &  -2.058  &  5.1  & 0.82   &     0.14   &    0.69   &     3.45      &  1.3     &    7.8     \\
         &   Red &  190.512 &  -2.050  &   5.0  & 0.73  &     0.08   &    0.40   &     1.97      &   1.1    &     5.4    \\
W-10     &   Blue &  190.529 &  -2.358  &  4.2  & 0.55   &     0.04   &    0.18   &     0.75      &  1.0     &    2.3     \\
W-11     &   Red &  191.948 &  -2.520  &   2.1  & 0.82  &     0.11   &    0.24   &     0.50      &   2.7    &     0.6    \\
W-12     &   Blue &  192.450 &  -2.729  &  2.7  & 0.82   &     0.11   &    0.28   &     0.74      &  2.3     &    1.1     \\
         &   Red &  192.504 &  -2.721  &   1.8  & 0.80  &     0.14   &    0.24   &     0.41      &   2.7    &     0.5    \\
W-13     &   Red &  192.721 &  -2.538  &   2.7  & 0.86   &    0.17  &     0.45   &     1.21      &   2.1     &    1.8     \\
W-14     &   Red &  193.008 &  -2.461  &   4.2  & 0.45   &     0.08   &    0.33   &     1.36      &  0.7     &     5.8     \\
W-15     &   Red &  193.021 &  -2.575  &   2.2  & 0.85   &     0.11   &    0.24   &     0.52      &  2.5     &     0.6     \\
W-16     &   Blue &  193.592 &  -3.075  &  1.9  & 0.44   &     0.03   &    0.05   &     0.09      &  1.8     &     0.1     \\
W-17     &   Blue &  193.985 &  -3.380  &  3.4  & 0.79   &     0.09   &    0.31   &     1.04      &  1.8     &     1.8     \\
W-18     &   Blue &  194.083 &  -2.880  &  2.0  & 0.76   &     0.10   &    0.20   &     0.40      &  3.0     &     0.4     \\
W-19     &   Blue &  194.179 &  -2.846  &  2.6  & 0.87   &     0.11   &    0.30   &     0.77      &  2.8     &     0.9     \\
         &   Red &  194.187 &  -2.827  &   1.9  & 0.56   &     0.14   &    0.27   &     0.51      &  1.8     &     0.9     \\
\enddata
\end{deluxetable}

\end{document}